\documentclass[10pt,journal,compsoc]{IEEEtran}

\usepackage[T1]{fontenc}
\usepackage{booktabs} 
\usepackage{graphicx}
\usepackage{float}
\usepackage{color}
\usepackage{array}
\usepackage{amssymb}
\usepackage{amsmath}
\usepackage{url}
\usepackage{subfig}
\usepackage{multirow}
\usepackage{hyperref}
\usepackage{pifont}
\usepackage{enumitem}
\usepackage{makecell} 
\usepackage[table]{xcolor}
\usepackage{colortbl}

\ifCLASSOPTIONcompsoc
  \usepackage[nocompress]{cite}
\else
  \usepackage{cite}
\fi
\ifCLASSINFOpdf
\else
\fi
\usepackage{makecell}

\newcommand{\reffig}[1]{{Fig.~\ref{#1}}}
\newcommand{\reftab}[1]{{Tab.~\ref{#1}}}
\newcommand{\refsec}[1]{{Sec.~\ref{sec:#1}}}
\newcommand{\refeq}[1]{{Eq.~\ref{#1}}}

\begin{document}\sloppy

%
\title{Think2Sing: Orchestrating Structured Motion Subtitles for Singing-Driven 3D Head Animation}

\author{
 	Zikai~Huang,
  Yihan~Zhou,
 	Xuemiao~Xu,
 	Cheng~Xu,
  Xiaofen~Xing,~\IEEEmembership{Member,~IEEE}, \\
  Jing~Qin,~\IEEEmembership{Senior Member,~IEEE},
 	and~Shengfeng He,~\IEEEmembership{Senior Member,~IEEE}
     \IEEEcompsocitemizethanks{
     \IEEEcompsocthanksitem \textup{Zikai Huang, Yihan Zhou, and Xuemiao Xu are with the School of Computer Science and Engineering, South China University of Technology, Guangdong, China. Xuemiao Xu is also with Guangdong Engineering Center for Large Model and GenAI Technology, Guangdong Provincial Key Lab of Computational Intelligence and Cyberspace Information. E-mail: 202210188523@mail.scut.edu.cn; yhzhou.ai@outlook.com; xuemx@scut.edu.cn.}
     \IEEEcompsocthanksitem \textup{Cheng Xu and Jing Qin are with the Centre for Smart Health, Hong Kong Polytechnic University, Hong Kong, and are also with the CAS-Hong Kong Joint Laboratory for Multimodal Medical Molecular Imaging. Email: cschengxu@gmail.com; harry.qin@polyu.edu.hk.}
     \IEEEcompsocthanksitem \textup{Xiaofen Xing is with the School of Electronic and Information Engineering, South China University of Technology, Guangdong, China, and also with Pazhou Lab, Guangzhou, Guangdong, China. Email: xfxing@scut.edu.cn.}
     \IEEEcompsocthanksitem \textup{Shengfeng He is with the School of Computing and Information Systems, Singapore Management University. Email: shengfenghe@smu.edu.sg.}
     }
 }
\markboth{IEEE Transactions on Visualization and Computer Graphics}%
{Huang \MakeLowercase{\textit{et al.}}: Think2Sing}
%




\makeatletter
\long\def\@IEEEtitleabstractindextextbox#1{\parbox{0.922\textwidth}{#1}}
\makeatother

\IEEEtitleabstractindextext{%
\begin{abstract}
Singing-driven 3D head animation is a compelling yet underexplored task with broad applications in virtual avatars, entertainment, and education.
Compared to speech, singing conveys richer emotional nuance, dynamic prosody, and lyric-conditioned semantics, necessitating the synthesis of fine-grained and temporally coherent facial motion.
Existing speech-driven approaches, which typically map audio directly to motion through implicit phoneme-to-viseme correspondences, often yield over-smoothed, emotionally flat, and semantically inconsistent results.
These limitations render them inadequate for the unique demands of singing-driven animation.
To address this challenge, we propose \textit{Think2Sing}, a unified diffusion-based framework that integrates pretrained large language models to generate semantically consistent and temporally coherent 3D head animations conditioned on both lyrics and acoustics.
Central to our approach is the introduction of an auxiliary semantic representation called \textit{motion subtitles}, derived via a novel Singing Chain-of-Thought reasoning process augmented with acoustic-guided retrieval.
These subtitles contain precise timestamps and region-specific motion descriptions, serving as interpretable and expressive motion priors that guide the animation process.
Rather than learning a direct audio-to-motion mapping, we reformulate the task as a \textit{motion intensity prediction} problem, which quantifies the dynamic behavior of key facial regions.
This reformulation decomposes the complex mapping into tractable subtasks, facilitates region-wise control, and improves the modeling of subtle and expressive motion patterns.
To support this task, we construct the first multimodal singing dataset comprising synchronized video clips, acoustic descriptors, and structured motion subtitles.
This dataset enables expressive and diverse motion learning under rich acoustic and semantic conditioning.
Extensive experiments demonstrate that Think2Sing significantly outperforms state-of-the-art methods in realism, expressiveness, and emotional fidelity.
Furthermore, our framework supports flexible subtitle-conditioned editing, enabling precise and user-controllable animation synthesis.
Visual results are available on the \href{https://zikaihuangscut.github.io/Think2Sing/}{\textcolor{blue}{project page}}.
\end{abstract} 

}


\maketitle

\IEEEdisplaynontitleabstractindextext

%
\IEEEpeerreviewmaketitle

\section{Introduction}
Singing-driven 3D head animation has recently emerged as a promising research direction with broad applications in virtual and augmented reality, digital entertainment, and education~\cite{yu20193d,wu2023singinghead,liu2024musicface,xie2025let}. Compared to speech, singing exhibits greater dynamism, emotional richness, and expressive variability~\cite{quinto2013emotional,livingstone2013acoustic,eyben2015emotion,livingstone2015common}. Performances often involve dramatic fluctuations in pitch, rhythm, and intensity, accompanied by nuanced facial expressions that convey subtle affective states. These characteristics introduce unique challenges for animation, demanding temporally coherent and semantically aligned synthesis that captures both prosodic features and lyrical intent. On the other hand, unlike video-based portrait animation~\cite{deng2025occlusion,li2024singer,xie2024x}, which modifies pixel-level appearances, 3D head animation explicitly models facial geometry and motion, allowing for flexible, expressive, and controllable animation.

Despite recent progress in speech-driven facial animation~\cite{fan2022faceformer,xing2023codetalker,stan2023facediffuser,peng2023selftalk}, singing-driven head animation remains significantly underexplored. Existing methods typically learn a direct mapping from audio to motion based on phoneme-to-viseme correspondences. While effective for speech, these approaches often produce over-smoothed and emotionally flat results when applied to singing. More critically, they lack mechanisms to model higher-level semantic structures or align motion with lyrical content, leading to animations that are both visually bland and semantically disconnected. Furthermore, current singing datasets are limited in scope: they often provide only audio input, lacking explicit textual or structural annotations that describe fine-grained facial and head movements. The scarcity of richly annotated motion data hinders the model's ability to learn expressive and diverse motion patterns grounded in both acoustic and semantic cues.

However, singing inherently exhibits structured and expressive motion patterns that are temporally coupled with both audio and lyrics~\cite{quinto2013emotional,quinto2014singing,eyben2015emotion,livingstone2015common}. Performers modulate emotional intensity and facial expressiveness dynamically across musical phrases, rather than remaining in fixed emotional states. This observation motivates a reformulation of the problem: instead of directly regressing from acoustics to motion trajectories, we introduce a structured reasoning process that infers explicit, time-aligned motion guidance.

We term this guidance \textit{motion subtitles}: a SubRip Subtitle (SRT)-style representation that encodes region-specific motion descriptions with precise timestamps. These subtitles capture the expressive dynamics of facial regions and serve as interpretable priors for downstream motion synthesis. By modeling these subtitles as an intermediate representation, we enable stronger semantic alignment and better prosodic understanding. Inspired by recent advances in large language models~(LLMs), we further hypothesize that LLMs, with their powerful reasoning capabilities, can effectively infer motion subtitles from lyrics and acoustic cues.

Building on these insights, we propose \textbf{Think2Sing}, an LLM-assisted, diffusion-based framework for expressive singing-driven 3D head animation. Our method introduces a singing-specific chain-of-thought reasoning strategy, augmented with an acoustic-guided retrieval module, to generate temporally coherent and emotionally adaptive motion subtitles conditioned on lyrics and audio. These subtitles provide explicit, fine-grained, and time-aligned motion instructions that guide the generation of semantically consistent and emotionally expressive facial dynamics.
To enhance control and expressiveness, we further propose a region-specific \textit{motion intensity} representation that quantifies dynamic activity across facial regions.In contrast to vertex-based or FLAME-based approaches, which often suffer from over-smoothing or lack fine-grained controllability, our representation preserves regional independence and enables accurate modeling of subtle expressions. This design complements our framework by allowing motion subtitles to be mapped directly to corresponding intensity heads, reducing learning ambiguity and improving synthesis quality.

To support this framework, we construct \emph{SingMoSub}, the first multimodal dataset for singing-driven 3D head animation, comprising synchronized singing clips with detailed acoustic descriptors and region-wise motion subtitles. These annotations span key facial regions, including eyebrows, eyes, mouth, and neck pose, enabling interpretable supervision for both semantic and prosodic modeling. Extensive experiments demonstrate that Think2Sing outperforms existing methods in realism, expressiveness, and emotional fidelity. In addition to high-quality generation, our framework supports subtitle-conditioned editing, allowing for flexible, interpretable, and user-controllable animation synthesis.

The main contributions of this work are fourfold:
\begin{itemize}[leftmargin=*]
    \item We propose Think2Sing, a unified framework that leverages LLMs to infer structured motion subtitles for expressive 3D head animation driven by singing. To the best of our knowledge, this is the first approach that utilizes motion priors derived from LLMs to produce time-aligned, semantically grounded, and emotionally consistent animations synchronized with singing performances.
    \item We develop a novel Singing Chain-of-Thought (Sing-CoT) reasoning scheme, enhanced by an Acoustic-Guided Retrieval-Augmented (AGRA) strategy, to generate motion subtitles with precise timestamps and lyric-conditioned, region-specific motion cues.
    \item We introduce a motion intensity representation that preserves regional independence, enables fine-grained control, and improves the accuracy and expressiveness of motion synthesis.
    \item We curate SingMoSub, the first multimodal dataset for singing-driven 3D head animation, featuring synchronized clips with detailed motion subtitles and acoustic descriptors for learning realistic and emotionally rich facial dynamics.
\end{itemize}

\section{Related Work}
\subsection{Audio-Driven 3D Head Animation}
Early studies~\cite{edwards2016jali,taylor2017deep,taylor2012dynamic,xu2013practical,zhou2018visemenet,vocaset,fan2022faceformer,xing2023codetalker,peng2023selftalk,wu2023speech,richard2021meshtalk} on audio-driven facial animation mainly addressed lip synchronization, mapping speech audio to mouth movements to achieve accurate phoneme-viseme alignment.
While effective for speech, these approaches capture only limited facial dynamics and fail to represent the prosodic, emotional, and semantic richness required for singing.
More recent works~\cite{karras2017audio,danvevcek2023emotional,peng2023emotalk,wang2020mead,song2024expressive,liu2024emoface,nocentini2025emovoca} extended the scope to full 3D head motion and expressive facial animation.
For example, EMOTE~\cite{danvevcek2023emotional} generates emotional 3D head motions using audio and predefined emotion labels.
EmoTalk~\cite{peng2023emotalk} disentangles emotion and linguistic content into separate latent spaces, enabling emotion control while preserving content.
However, these approaches treat emotion and audio as fully decoupled and rely heavily on predefined emotion labels or intensity parameters to control facial expressions, and thus cannot automatically infer dynamic emotional states and intensity directly from the audio.
Recently, researchers have explored singing-driven head animation, which requires more expressive and nuanced head movements than speech~\cite{liu2024musicface,wu2023singinghead}.
However, methods such as SingingHead~\cite{wu2023singinghead} and MusicFace~\cite{liu2024musicface} rely solely on audio's ASR features as driving conditions, limiting their ability to capture the full complexity of singing prosody and semantics.
Moreover, treating emotion and content as independent signals is inappropriate in singing, where expressive intent and head motion are inherently intertwined.
Differently, our method leverages the strong reasoning capabilities of large language models~(LLMs) to infer fine-grained, semantically rich facial motion subtitles directly conditioned on lyrics and acoustics. 
These subtitles provide interpretable guidance beyond raw audio features, enabling expressive, lyricaware animation synthesis without relying on predefined emotionlabels or costly manual annotations. 

\subsection{Text-guided 3D Human Motion Generation}
Text-guided 3D human motion generation aims to generate realistic human motion sequences from natural language descriptions.
Existing methods can be generally categorized into body motion generation and head motion generation.
While text-guided body motion generation has witnessed substantial progress, head motion generation still remains underexplored.
In body motion generation, the goal is to synthesize actions consistent with textual semantics, often without strict temporal constraints.
Existing approaches include recurrent neural networks~(RNNs)\cite{wang2020learning,guo2020action2motion,martinez2017human}, transformers\cite{petrovich2021action,petrovich2022temos,gong2023tm2d}, and diffusion-based models~\cite{tevet2022human,shafir2023human,chen2024text,zhang2024motiondiffuse}.
In contrast, head motion generation typically relies on coarse-grained textual descriptions, such as emotion labels or simple temporal adverbial phrases~\cite{ma2023talkclip,sun2024avi,wu2024mmhead}, which are insufficient for modeling the fine-grained, dynamic expressions required in singing.
Among the existing works, MMHead~\cite{wu2024mmhead} is most relevant to our work, which introduces a multi-modal framework combining both audio and text to generate 3D facial animation.
However, the its annotations are limited in temporal cues, as they rely on coarse temporal adverbial phrases rather than precise motions intervals, limiting the granularity of synthesis.
In contrast, our method introduce motion subtitles as explicit textual guidance, which encapsulate both semantic context and precise timestamps, enabling more accurate and fine-grained motion modeling.
More importantly, unlike prior works that require user-provided textual input, our subtitles are automatically inferred from singing audio through an LLM-assisted Singing Chain-of-Thought scheme with acoustic-guided retrieval augmentation, yielding temporally precise and semantically rich annotations.
This fully automatic design leads to more expressive and temporally coherent motion synthesis compared to existing methods.

\subsection{3D Head Animation Datasets for Singing} 
Despite increasing interest in singing-driven facial animation, expressive 3D head motion datasets tailored for singing remain scarce and limited in both scale and quality.
Several datasets have been introduced. For example, the RAVDESS dataset~\cite{ravdess} includes a small subset of singing recordings, but clips are restricted to isolated single sentences with artificially separated lyrical and emotional content, failing to capture the continuous, emotionally integrated nature of real singing.
This structure fails to capture the continuous and emotionally integrated nature of authentic singing performances.
Song2Face~\cite{iwase2020song2face} provides around 2 hours of recordings from 7 singers and therefore suffer form limited diversity in vocal style and facial expressiveness.
MusicFace~\cite{liu2024musicface} includes karaoke-style performances from 6 subjects, which restricts natural expressive variation due to the imitation-based setup.
More recently, SingingHead~\cite{wu2023singinghead} presents a larger dataset with 27 hours of recordings from 76 singers, but audio quality is degraded by background noise, and a large portion (41\%) comes from amateurs whose facial dynamics are less expressive and natural than professionals.
To combat these limitations, we introduce the first large-scale multi-modal 3D singing head animation dataset, which contains synchronized singing clips, acoustic descriptors, and fine-grained, time-aligned head motion subtitles.
These rich annotations offer detailed supervision for modeling prosody, semantics, and expressiveness, enabling the development of more realistic and lyric-aware head animation models.

\begin{figure*}[t]
\centering
\includegraphics[width=\linewidth,scale=1.00]{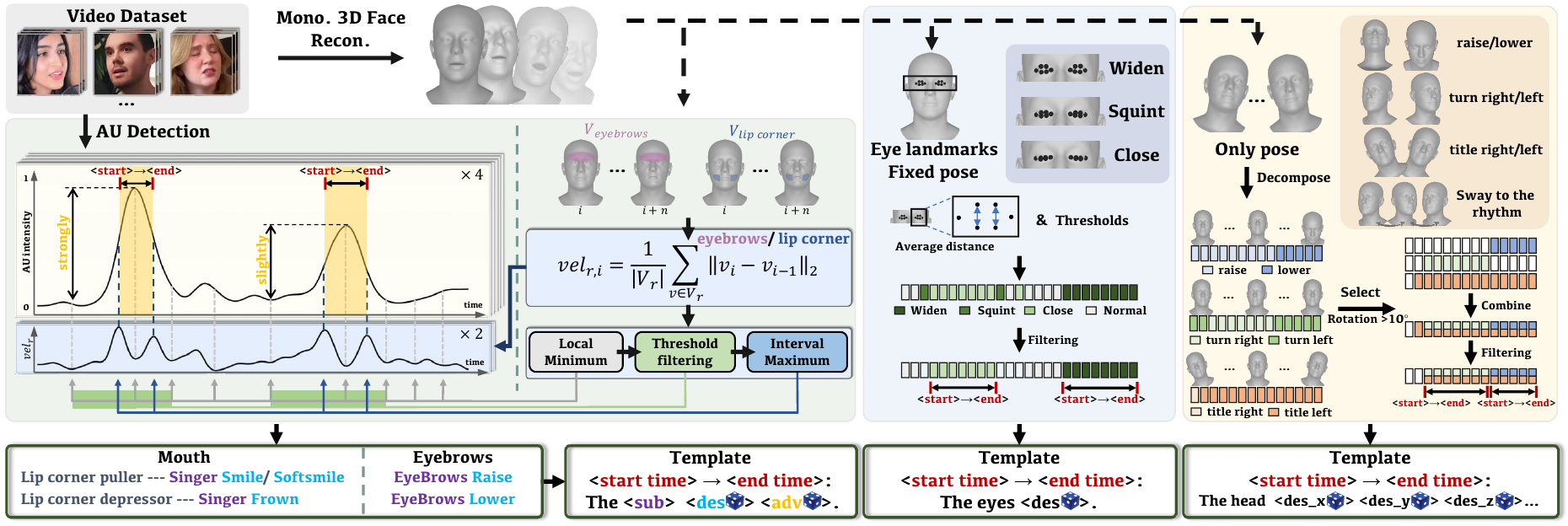}
\caption{
\textbf{The pipeline of motion subtitle annotation.}
We present an automated pipeline that efficiently generates diverse, fine-grained motion subtitles with precise timestamps. Together with the corresponding video frames, these subtitles constitute the first dataset that delivers region-wise, time-synchronized annotations for singing scenarios, enabling more accurate and fine-grained modeling of singing-motion relationships in expressive, singing-driven 3D head animation.
}
\vspace{-10pt}
\label{motion_sub_annotation}	
\end{figure*}

\section{SingMoSub Dataset}
We curated a large-scale dataset, termed SingMoSub, containing over 37 hours of singing videos sourced from platforms such as YouTube, BiliBili, and other publicly available repositories.
For each frame, we extract FLAME parameters~\cite{flame} using the state-of-the-art monocular 3D face reconstruction method EMOCA v2~\cite{danvevcek2022emoca, feng2021learning, filntisis2022visual} to represent 3D head motion.
The dataset prioritizes fine-grained motion annotation across both temporal and spatial dimensions, including time-aligned region-wise motion subtitles and acoustic descriptions.
To the best of our knowledge, it is the first dataset to include motion subtitles tailored for singing scenarios, providing temporally aligned, region-wise annotations of 3D head motion together with acoustic descriptions.
For more implementation details, please refer to the supplementary material.

\subsection{Data Annotation}
\subsubsection{Motion Subtitle}
To capture dynamic and expressive motions unique to singing, we propose \emph{motion subtitle}, a structured textual representation that describes the dynamic motion of a specific head region over a defined time interval, providing both fine-grained temporal precision and semantic guidance to capture subtle variations in motion.
Each motion subtitle follows the SRT-style format: \texttt{``<start\_time> --> <end\_time>: <region> <description>''}, ensuring both clarity and consistency across the dataset.
Unlike prior datasets that rely on static, plain-text descriptions without temporal alignment, our annotations are both temporally localized and semantically rich.
To this end, we design an annotation protocol for the proposed motion subtitles (\reffig{motion_sub_annotation}) that generates time-aligned motion subtitles across four key regions: eyebrows, mouth, eyes, and neck pose.

\textbf{Eyebrows and Mouth.}
Eyebrow and mouth movements play a central role in conveying emotions and lyrical delivery.
To localize expressive intervals with high precision, we combine framewise Action Unit~(AU)~\cite{ekman1978facial} normalized intensity with vertex-level dynamics.
A region-wise velocity curve is computed from frame-to-frame vertex displacements under a fixed neck pose.
Local minima and maxima in this curve identify temporal transitions and the moments of peak motion.
Expression intervals are identified by pairing successive minima whose AU intensity difference exceeds a predefined threshold, thereby filtering out insignificant changes.
For each interval, the frame with the maximum velocity is selected as the onset, and the offset is determined by the subsequent validated interval.
This procedure achieves both temporal precision and semantic validity, with velocity extrema offering robust temporal boundaries and AU intensities confirming meaningful expressive variations.
The \texttt{description} is constructed using the following template: \texttt{``<motion> <intensity>''}, where \texttt{<motion>} and \texttt{<intensity>} are selected from the corresponding predefined vocabulary to ensure diversity and consistency in the descriptions.

\textbf{Eyes.}
During singing, eye movements often reflect expressive nuances, such as widening to emphasize intensity or squinting to suggest subtle emotions.
To capture these dynamics, we measure eyelid openness relative to neutral FLAME parameters and annotate eye states, including widening, squinting, and closing, which align with the expressive patterns typically observed in singing performance.
The \texttt{description} for eyes is constructed using the following template: \texttt{``The eyes <state>.''}, where $\texttt{<state>} \in \{ \texttt{widen}, \texttt{squint}, \texttt{close}\}$ are sampled from predefined vocabulary.

\textbf{Neck Pose.}
Head movements are a crucial component of expressive singing, contributing both to the emotion conveyance and overall performance.
We annotate head movements along three axes: vertical, horizontal, and lateral.
Rotations below a predefined threshold are discarded to filter out insignificant movements.
To reduce redundancy, frequent alternating left-right turns are generalized as ``sway to the rhythm''.
Neck pose annotations follow the template: \texttt{``The head <des\_x>, <des\_y>, <des\_z>''}, where each descriptor specifies motion along the corresponding axis and is selected from a predefined vocabulary, if present.

\subsubsection{Acoustic Description}
Several studies~\cite{hakanpaa2019emotion, scherer2017expression} have shown that identical lyrics can express very different emotions depending on vocal delivery.
For instance, \textit{``I'm fine''} may sound sad when sung softly at a low pitch, but confident when delivered with greater volume and higher pitch.
To fully exploit acoustic features for modeling more accurate audio-motion relationships in singing, drawing inspiration from~\cite{wu2024speechcuellm}, we employ an acoustic analysis approach.
For each lyric line, we extract three key acoustic features, namely, volume, pitch, and singing rate from the singing vocal, which together form the basis of the acoustic description.
For each singer, we compute statistical characteristics of volume and pitch, and use the 25th and 75th percentiles to define three discrete levels: low, moderate, and high.
During inference, global thresholds derived from the training set are applied to normalize and classify features for unseen singers.
The acoustic description annotations are used for the retrieval grounding (\refsec{AGRA}).

\subsection{Comparison with Existing Datasets}
We compare our dataset with existing datasets in~\reftab{dataset_comparison}.
SingingHead~\cite{wu2023singinghead} is currently the largest and most widely used 3D singing head animation dataset, spanning 27.10 hours.
MMHead~\cite{wu2024mmhead}, a more recent contribution, provides fine-grained motion descriptions for 3D head animation.
Given their direct relevance, both datasets serve as key baselines for evaluating our dataset.

\begin{table}[t]
\noindent\begin{minipage}{\linewidth}
    \centering
    \captionof{table}{
    \textbf{Comparison with existing datasets.}
    Silence Rate is reported only for the singing-oriented datasets.
    }
    \setlength{\tabcolsep}{2pt}
    {
        \renewcommand\arraystretch{1.2}
        \begin{tabular}{c|ccccc} \hline
            Dataset                                     & Duration      & \makecell{Silence\\Rate} & \makecell{Singing\\Oriented} & \makecell{Motion\\Subtitle} & \makecell{Acoustic\\Description} \\ \hline
            MMHead~\cite{wu2024mmhead}                  & 49.00 h       & -                        & \ding{55}                     & \ding{55}                    & \ding{55}                          \\
            SingingHead~\cite{wu2023singinghead}        & 27.10 h       & 16.19\%                  & \ding{51}                     & \ding{55}                    & \ding{55}                          \\ \hline
            \rowcolor{gray!20}
            SingMoSub (Ours)                            & 37.13 h       & 1.79\%                   & \ding{51}                     & \ding{51}                    & \ding{51}                          \\ \hline
        \end{tabular}
        \label{dataset_comparison}
    }
\end{minipage}
\end{table}

To better evaluate the datasets from a singing perspective, we introduce \textit{Silence Rate}, a metric defined as the ratio of non-vocal to total segments.
A higher Silence Rate indicates a greater proportion of non-vocal content, which provides limited training and evaluation value.
As shown in~\reftab{dataset_comparison}, our dataset achieves a substantially lower Silence Rate~(1.79\% vs. 16.19\% for SingingHead), highlighting its stronger focus on vocal activity and greater utility for voice-driven motion modeling.

\begin{figure}[t]
    \centering
    \includegraphics[width=.4\textwidth]{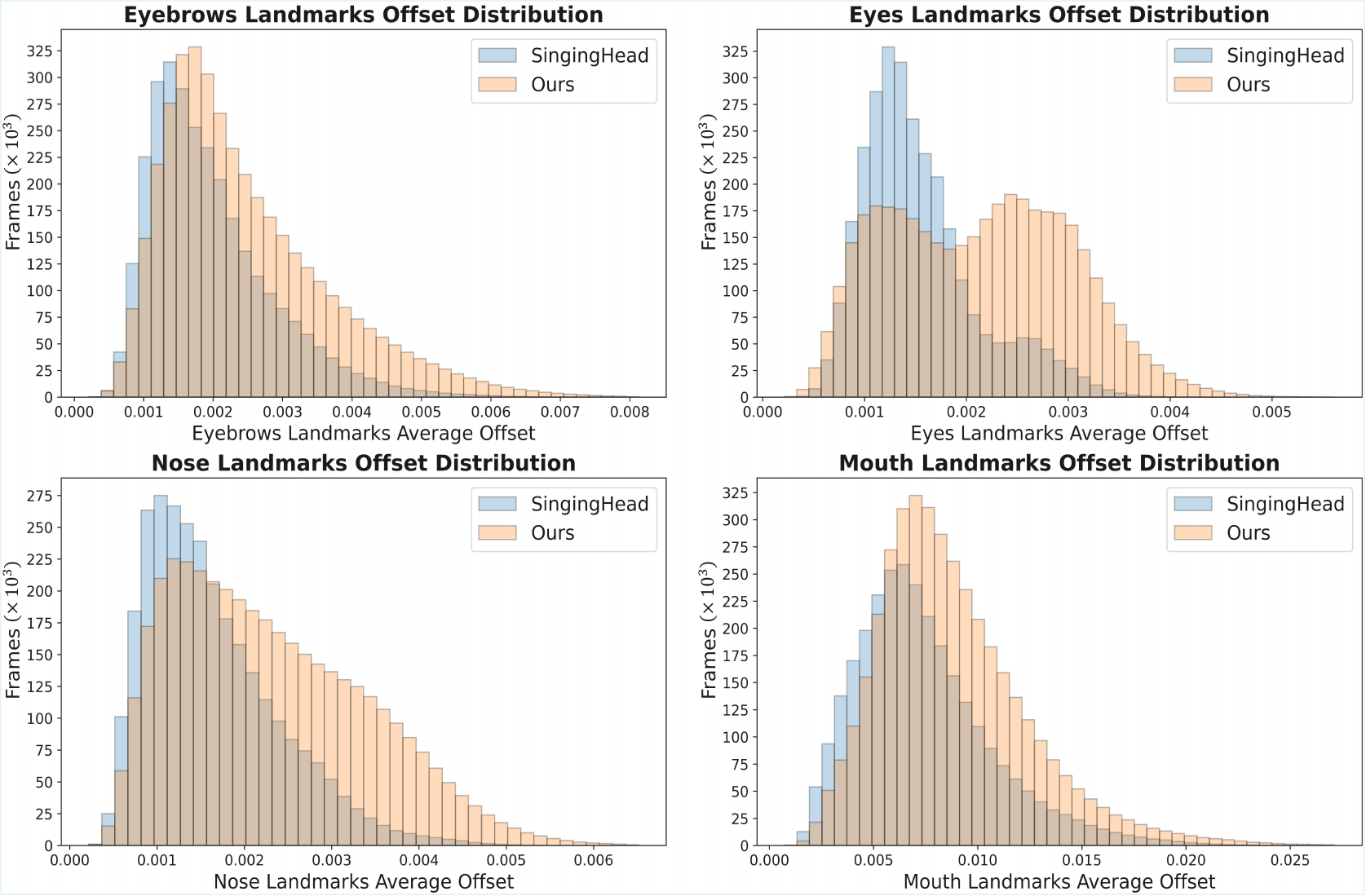}
    \caption{
        \textbf{Visualization of region-wise landmark offset distribution.}
        Our dataset demonstrates a larger dynamic range across all regions compared to the SingingHead dataset.
        } 
    \label{landmarks_comparison}	
  \end{figure}

\begin{figure}[t]
    \centering
    \includegraphics[width=.45\textwidth]{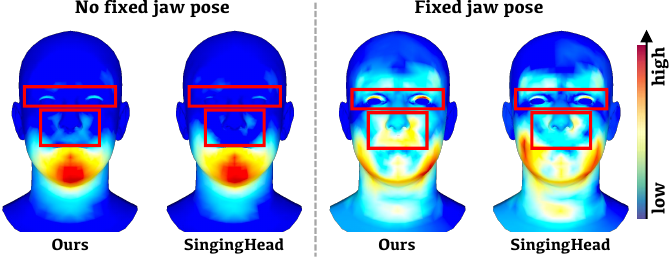}
    \caption{
        \textbf{Visualization of vertex displacement in our dataset compared to SingingHead.}
        All columns are shown using a unified scale.
        } 
    \label{heatmap}	
  \end{figure}

Unlike speech-driven head animation, singing involves richer and more diverse facial movements that extend beyond the mouth region.
To quantify this expressiveness, we measure vertex displacement relative to a neutral face across multiple regions, including the eyebrows, eyes, nose, and mouth.
To evaluate this, we calculate the average vertex displacement across all frames for multiple facial regions, including the eyebrows, eyes, nose, and mouth.
As illustrated in \reffig{landmarks_comparison}, our dataset consistently exhibits larger displacement than SingingHead, indicating enhanced facial expressiveness and dynamic variability.
To further isolate the role of singing content in mouth movements, we visualize vertex displacement heatmaps under fixed and unfixed jaw pose conditions (\reffig{heatmap}).
This results show that, unlike SingingHead, our dataset captures more pronounced motions across full face, demonstrating a broader and more expressive range of facial dynamics in singing scenarios.

Building upon the demonstrated richness of facial dynamics, another key strength of our dataset lies in its comprehensive annotation scheme.
Unlike MMHead, which provides only clip-level descriptions without temporal specificity, our dataset introduces region-wise motion subtitles the preceise timestamps.
These subtitles are formatted in a region-specific SRT format, enabling fine-grained temporal alignment.
In addition, our dataset is the first 3D head animation resource to include detailed acoustic description annotations.
This multimodal design not only captures subtle facial movements over time but also enables a deeper analysis of the interplay between vocal acoustics and facial motion in singing.
Collectively, these features distinguish our dataset from existing resources and establish it as a superior benchmark for modeling expressive singing performances.
We will release the dataset publicly upon acceptance to support further research in expressive 3D head animation.

\section{Think2Sing}
Given a singing audio sequence $A_{1:L}$, our goal is to synthesize a sequence of 3D head motion $M_{1:L} = \{m_1, \cdots, m_L\}$ that is temporally aligned with the audio and captures the expressive dynamics of singing, where $m_i = [\psi_i, \theta^{neck}_i, \theta^{jaw}_i]$ denoted the FLAME~\cite{flame} parameters at frame $i$.
To this end, we first introduce a Singing Chain-of-Thought~(\refsec{Sing-CoT}) reasoning scheme with Acoustic-Guided Retrieval Augmentation~(\refsec{AGRA}) to generate fine-grained, region-specific motion subtitles from time-aligned lyrics and acoustic features.
These motion subtitles provide explicit and precise guidance for estimating motion intensity~(\refsec{motion_intensity}), an intermediate proxy that captures the dynamics of key facial regions while maintaining disentangled control.
A Subtitle-guided Proxy-based Motion Generator~(\refsec{motion_generator}) then leverages these subtitles to produce neck pose and facial motion intensities, which are subsequently converted into the final FLAME head parameters.

\begin{figure*}[t]
\centering
\includegraphics[width=\linewidth,scale=1.00]{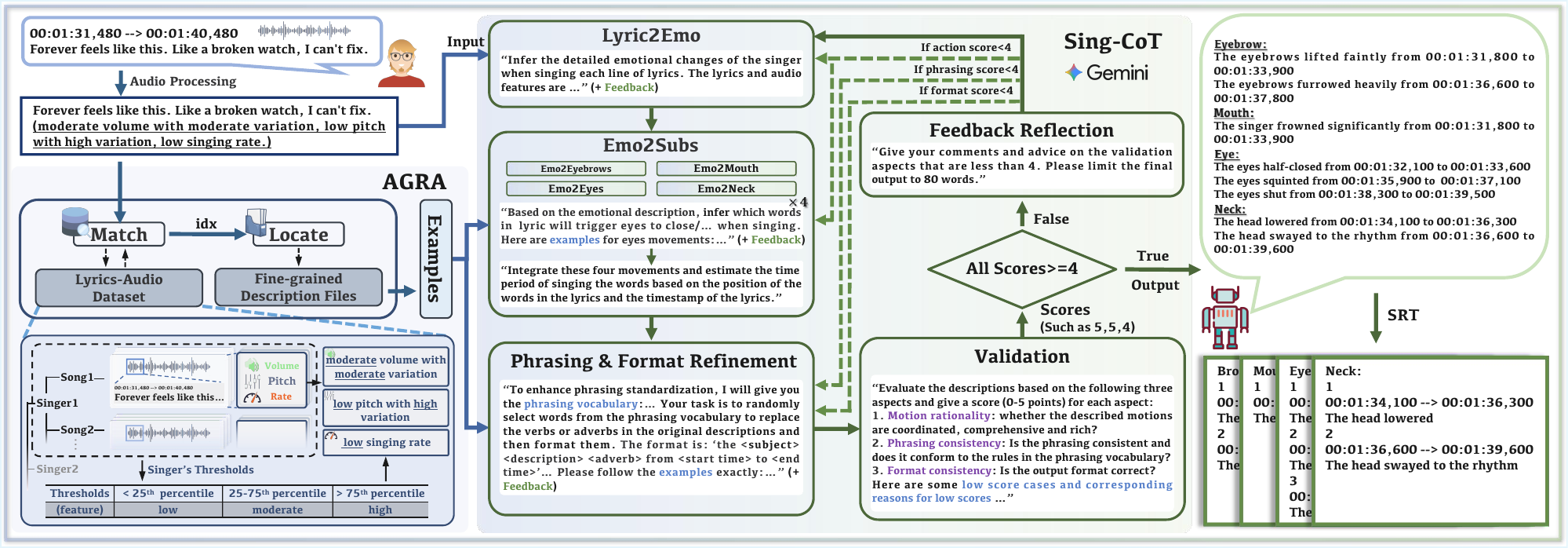}
\caption{
\textbf{The pipeline of expressive singing motion subtitles generation.}
We first introduce Acoustic-Guided Retrieval Augmentation~(AGRA), a retrieval-augmented generation framework that incorporates acoustic descriptors to retrieve semantically prosodically aligned examples.
Based on these examples, an emotion-aware Singing Chain-of-Thought reasoning strategy is further proposed for generating temporally aligned, fine-grained, and region-wise motion subtitles.
} 
\vspace{-10pt}
\label{agra_singcot}	
\end{figure*}

\subsection{Singing Motion Subtitle Generation}
Generating expressive facial animations for singing requires fine-grained, temporally aligned, and emotionally grounded motion control.
Previous approaches based solely on audio or static transcripts often fail to capture the nuanced dynamics of lyrical expression.
To address this, we propose a novel module for expressive motion subtitle generation that leverages LLM with retrieval-augmented generation~(RAG)~\cite{lewis2020retrieval} and Chain of Thought~(CoT)~\cite{wei2022chain} reasoning illustrated in \reffig{agra_singcot}.
Inspired by principles in facial motion analysis and affective computing~\cite{ekman1978facial, cohn2007observer}, we empirically decompose head animation into four functional regions: eyebrows, eyes, mouth, and neck pose.
This partitioning reflects the distinct communicative roles of different regions in conveying affect, attention, and articulation.
By extracting time-aligned lyrics and acoustic descriptors from the input singing audio, the proposed module produces part-aware, temporally synchronized motion descriptions formatted as SubRip Subtitle~(SRT) files.
Unlike static transcripts, SRT files contain both verbal content and precise temporal annotations, enabling tighter alignment between lyrical semantics and vocal dynamics.
This structure supports more accurate and expressive motion supervision.
The resulting motion subtitles serve as high-level semantic control signals, providing both temporal precision and expressive grounding for subsequent conditional facial animation generation.

\subsubsection{Acoustic Guided Retrieval Augmentation} \label{sec:AGRA}
To address the challenges of generating expressive facial motion from singing input, we propose an Acoustic-Guided Retrieval Augmentation (AGRA) Strategy for expressive grounding.
Sharing a similar spirit to RAG~\cite{lewis2020retrieval}, which enhances generative models through external knowledge retrieval, our proposed AGRA framework introduces a novel query formulation that synergistically combines textual and acoustic cues.
Unlike prior methods that rely solely on lyrics or limited contextual cues, AGRA constructs multimodal queries by jointly encoding timestamped lyrics and low-level acoustic descriptors derived directly from the input audio.
To capture comprehensive acoustic characteristics, we choose three representative descriptors, including volume, pitch, and singing rate, which are closely tied to facial expressiveness.
For instance, abrupt shifts in pitch, volume, or singing rate often correspond to distinctive facial movements.
This enriched query representation enables the retrieval of reference samples that are both semantically relevant and prosodically aligned with the target input.
These retrieved examples provide strong contextual grounding for LLM, guiding the generation of motion descriptions that are emotionally nuanced and temporally synchronized.

Specifically, given a singing audio $A_{1:L}$, we extract a set of time-aligned subtitle units in SRT format using the ASR model Whisper~\cite{radford2023robust}.
Each unit is represented as a tuple:
\begin{equation}
    sub_s = (\tau^{\text{start}}_s, \tau^{\text{end}}_s, [\texttt{lyric}_s, D_s]), \quad s \in \{1, \dots, n\},
\end{equation}
where $n$ is number of sentences detected, $\tau^{\text{start}}_s$ and $\tau^{\text{end}}_s$ denote the start and end timestamps, $\texttt{lyric}_s$ is the aligned lyric text.
Each $D_s$ consists of low-level prosodic descriptions extracted from the corresponding audio segment:
\begin{equation}
    D_s = \{\texttt{volume}_s, \texttt{pitch}_s, \texttt{rate}_s\},
\end{equation}
where $\texttt{volume}_s$, $\texttt{pitch}_s$, and $\texttt{rate}_s$ are the volume, pitch, and singing rate descriptions over the audio segment corresponding to $sub_i$.
The multimodal query $Q$ is defined as $Q = \{sub_s\}^n_{s=1}$.
Using the constructed multimodal query $Q$, we retrieve a set of $k$ relevant reference samples $R = \{sub^{\text{m}}_s\}_{s=1}^k$ from our curated dataset. Each retrieved sample $sub^{\text{m}}_s$ is represented as a tuple $(\tau^{\text{start}}_{r, s}, \tau^{\text{end}}_{r, s}, \texttt{motion}_{r,s})$, where $\tau^{\text{start}}_{r, s}$ and $\tau^{\text{end}}_{r, s}$ denote the start and end timestamps, and $\texttt{motion}_{r,s}$ denotes motion descriptions for region $r \in \{\text{eyebrows}, \text{eyes}, \text{mouth}, \text{neck}\}$.
With the acoustic descriptors incorporated into the retrieval process, the controllability and expressiveness of synthesized facial motions can be greatly enhanced.

\begin{figure*}[t]
\centering
\includegraphics[width=\linewidth,scale=1.00]{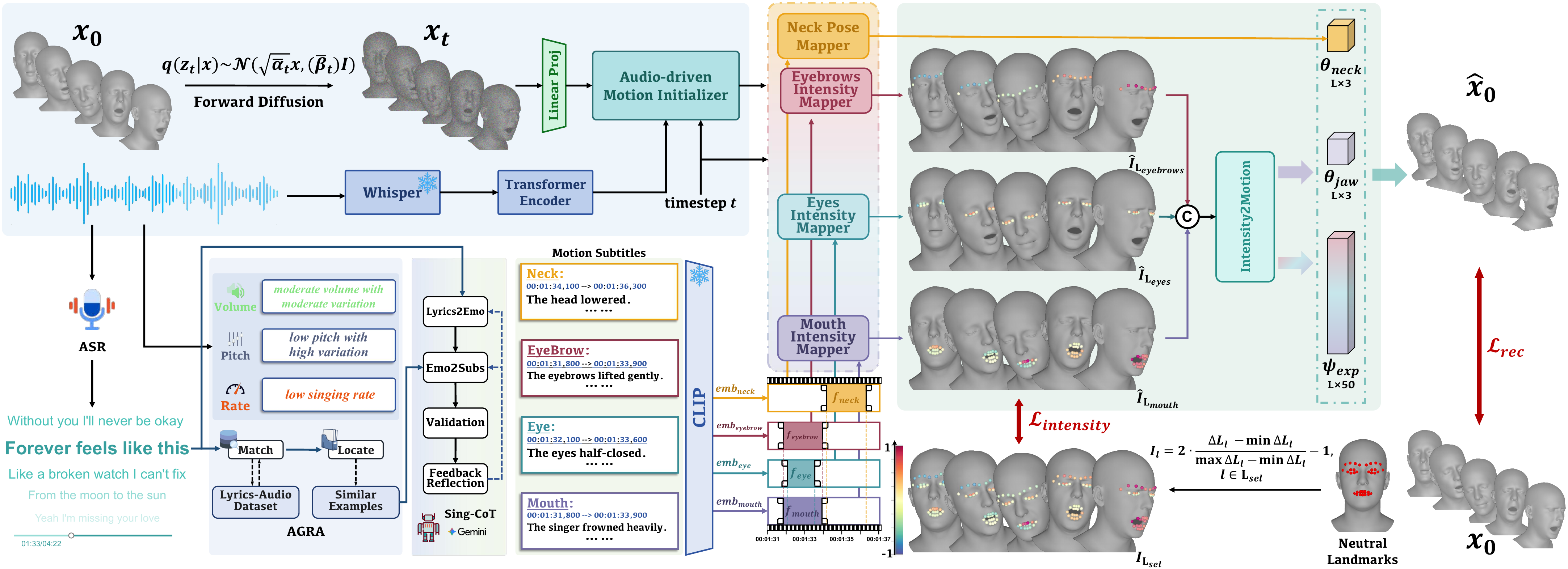}
\caption{
\textbf{Overview of our proposed Think2Sing.} 
Think2Sing is a unified diffusion-based framework conditioned on singing audio and inferred motion subtitles.
Audio features are first used to initialize a shared latent representation that encodes global prosody.
Motion subtitles, generated from time-aligned lyrics and acoustic descriptors via AGRA-Sing-CoT, are then incorporated as conditioning inputs.
By integrating the shared latent representation with region-specific subtitle features, the model predicts the neck pose and the motion intensities of the eyebrows, eyes, and mouth, where motion intensity $I_{\mathbf{L}_{sel}}$ serves as a proxy representation, effectively reducing learning complexity and enabling flexible region-wise control.
The Intensity2Motion predictor converts the estimated motion intensities into the final FLAME head parameters.
} 
\vspace{-10pt}
\label{overview}	
\end{figure*}

\subsubsection{Singing Chain of Thought for Subtitle Reasoning} \label{sec:Sing-CoT}
Directly generating structured motion subtitles with region-specific detail and precise temporal alignment is highly challenging, as the correspondence between acoustic/lyrical cues and facial motion is inherently non-deterministic.
Singing further exacerbates this dilemma by introducing multi-scale dynamics, encompassing rapid visemic events, phrase-level variations, and long-range emotional progressions, all of which must be synchronized coherently.
The spatial coupling of facial regions leads to additional complexity, while the scarcity of annotated datasets further limits reliable supervision.

To tackle this issue, we introduce a singing-specific emotion-aware reasoning scheme, Singing Chain of Thought~(Sing-CoT), which decomposes the generation process into a sequence of interpretable steps:

\emph{Step 1: Emotion extraction from lyrics.}
Expressive singing should faithfully reflect the emotional intent conveyed by the lyrics.
Establishing the underlying emotional tone provides a foundation for consistent and contextually appropriate motion planning.
Different emotional states tend to elicit distinct facial expressions.
For example, sadness may be reflected through lowered eyelids and furrowed brows, whereas joy often involves more frequent smiling and relaxed facial muscles.
We therefore first infer the dominant emotional cues from the lyrics, which guides subsequent reasoning and enables the generation of motion subtitles that are both semantically grounded and emotionally faithful.

\emph{Step 2: Acoustic-guided RAG motion subtitles generation.}
Using the AGRA module, relevant reference samples are retrieved based on these multimodal features.
The model then generates region-specific motion subtitles $\{sub_{r, s}\}$ conditioned on both the retrieved examples $R$ and the identified emotion $e$.

\emph{Step 3: Validation of generated subtitles.}
To ensure the accuracy and expressiveness of the generated motion subtitles, a validation mechanism assesses each generated result based on three criteria: physical plausibility, formatting correctness, and linguistic diversity.
In addition, curated negative examples are provided to the model as contrastive signals, helping to refine the quality assessment and filter out implausible or repetitive generations.

\emph{Step 4: Feedback reflection.}
If the validation score falls below a predefined threshold, Sing-CoT triggers an automatic regeneration process.
Targeted feedback is incorporated into the revised prompt to encourage the model to correct deficiencies, resulting in more coherent, expressive, and temporally precise outputs.

By jointly leveraging AGRA and Sing-CoT, we move beyond conventional audio-to-motion mapping and introduce a novel paradigm that produces temporally coherent, emotion-aware motion descriptions tailored to singing, thereby enabling controllable and expressive head animation.
We kindly refer readers to the appendix for detailed prompt template and implementation details.

\subsection{Subtitle-guided Proxy-Based Head Animation}
\subsubsection{Motion Intensity Proxy} \label{sec:motion_intensity}
We adopt FLAME parameters as the target representation for facial animation, given their widespread use in state-of-the-art monocular 3D face reconstruction and superior robustness in modeling large-amplitude head motions compared to direct vertex-based representations.
FLAME parameterizes facial dynamics through expression coefficients $\psi \in \mathbb{R}^{50}$, jaw pose $\theta^{\text{jaw}} \in \mathbb{R}^{3}$, and neck pose $\theta^{\text{neck}} \in \mathbb{R}^{3}$.
The combination of $\psi$ and $\theta^{\text{jaw}}$ controls local facial expressions, while $\theta^{\text{neck}}$ governs the global orientation and motion of the head.

Although FLAME is widely adopted, its expression basis tends to activate multiple muscle groups simultaneously, making it difficult to achieve fine grained control over individual facial components.
This coupling limits semantic interpretability from a human perceptual perspective.
To address this limitation, we introduce \emph{motion intensity}, a proxy representation that is spatially disentangled, semantically meaningful, and directly interpretable.
Unlike FLAME parameters, motion intensity captures local geometric variations in key facial regions over time, enabling finer control and  semantic interpretation.
This approach improves the modeling of subtle expressions and reduces ambiguity, facilitating more accurate and expressive motion synthesis.
Let $\mathbf{L}_i \in \mathbb{R}^{68 \times 3}$ denote the 3D facial landmarks at frame $i$, and $\mathbf{L}^{\text{neutral}} \in \mathbb{R}^{68 \times 3}$ be the landmarks corresponding to the neutral face. The amplitude of motion at landmark $l$ in frame $i$ is defined as:
\begin{equation}
\Delta L_{l, i} = \left \lVert \mathbf{L}_{l, i} - \mathbf{L}_l^{\text{neutral}} \right \rVert.
\end{equation}

To focus on regions most indicative of expressiveness, we define the selected landmark set:
\begin{equation}
\mathbf{L}_{\text{sel}} = \mathbf{L}_{\text{eyebrows}} \cup \mathbf{L}_{\text{eyes}} \cup \mathbf{L}_{\text{mouth}}.
\end{equation}

We normalize the motion amplitudes over the temporal window $[1, T]$ to obtain the motion intensity $I_{l, i}$ for each selected landmark:
\begin{equation}
I_{l, i} = 2 \cdot \frac{\Delta L_{l, i} - \min\limits_{i' \in [1, T]} \Delta L_{l, i'}}{\max\limits_{i' \in [1, T]} \Delta L_{l, i'} - \min\limits_{i' \in [1, T]} \Delta L_{l, i'}} - 1,\quad l \in \mathbf{L}_{\text{sel}}.
\end{equation}

The resulting intensity values lie in the range $[-1, 1]$, reflecting the relative magnitude of movement in each frame.
Compared to predicting dense vertex offsets across the entire facial mesh, our motion intensity representation offers a more compact, interpretable, and controllable alternative.
Modeling all vertex-level displacements is redundancy, overly sensitive to identity-specific geometry, and struggles to capture subtle motions beyond the mouth, often leading the model to regress toward a static average.
In contrast, by focusing on a subset of landmarks around the eyebrows, eyes, and mouth, motion intensity isolates the most emotionally salient dynamics, enhances temporal coherence, and enables precise, region-specific control.
This component-aware design isolates region-specific motion, enabling fine-grained control and semantic interpretability that are difficult to achieve with global mesh-based representations.
In addition, the motion intensity formulation models relative temporal variations instead of absolute geometric displacements from a fixed template.
This normalized representation emphasizes dynamic expressiveness over static deformation, improving temporal coherence and making it more robust to identity-specific geometry and inter-speaker variability.
By tracking frame-wise deviations from a neutral reference, the model better captures emotionally driven motion patterns across performances. Furthermore, the disentangled structure of motion intensity facilitates targeted supervision during training and flexible manipulation at inference time.
Collectively, these advantages represent a significant departure from traditional vertex-based regression and establish motion intensity as a semantically meaningful and effective control signal for expressive facial animation.

To convert motion intensity into actionable animation signals, we introduce an Intensity2Motion predictor, which learns to map sparse, component-aware intensity representations to FLAME expression coefficients $\psi$ and jaw pose $\theta^{jaw}$.
The Intensity2Motion predictor models a temporal mapping function $f_\phi$ parameterized by a lightweight Transformer, which captures the dynamic relationships between motion intensity patterns and the underlying facial motion representation:
\begin{equation}\label{eq:intensity2motion}
[\psi_{1:L}, \theta^{\text{jaw}}_{1:L}] = f_\phi(\mathbf{I}_{1:L}),
\end{equation}
where $\mathbf{I}_{1:L} = [I_{l, 1:L}], l \in \mathbf{L}_{\text{sel}}$ denotes the motion intensity sequence for the selected landmarks over time.
By establishing this correspondence, the Intensity2Motion predictor acts as a bridge between interpretable motion cues and the generative facial model, enabling accurate, expressive, and semantically grounded animation synthesis.

\subsubsection{Subtitle-guided Proxy-based Motion Generator}    \label{sec:motion_generator}
As illustrated in \reffig{overview}, our goal is to synthesize temporally coherent and expressive head animations conditioned on multimodal inputs, including singing features and motion subtitles.
To this end, we propose a Subtitle-guided Proxy-based Motion Generator for expressive head motion synthesis.
Let $\mathbf{x}_0 = M_{1:L}$ denote the target facial motion sequence.
The diffusion process consists of a forward noising phase and a reverse denoising phase.
The forward process is defined as:
\begin{equation}
    q(\mathbf{x}_t|\mathbf{x}_{t-1}) = \mathcal{N}(\mathbf{x}_t;\sqrt{1- \beta_t}\mathbf{x}_{t-1}, \beta_t\mathbf{I}).
\end{equation}
where $\beta_t$ is a pre-defined noise scheduler.
After $T$ steps, the original signal is transformed into a noise-like distribution, typically approaching a standard Gaussian $\mathcal{N}(\mathbf{0}, \mathbf{I})$.
The reverse process then attempts to recover the clean motion sequence $\mathbf{x}_0$ by gradually denoising $\mathbf{x}_T$ through a learned denoising network.
Following the DDPM formulation~\cite{ho2020denoising, tevet2022human}, we directly train the model to predict $\mathbf{x}_0$ given $\mathbf{x}_t$.

To capture the interdependence across different facial regions while preserving the capacity for region-specific control, we adopt a diffusion-based conditional generation scheme.
In the first stage, we extract audio features from the input singing audio using a pretrained Whisper encoder~\cite{radford2023robust}, which encodes acoustic cues.
These features are fed into an Audio-driven Motion Initializer that produces a shared latent representation encoding global prosodic information.

In the second stage, the shared latent vector $\mathbf{z}$ is routed to four dedicated mapper networks.
Each mapper generates a specific component of head animation conditioned on corresponding SRT-formatted motion subtitles.
Unlike static text prompts, each $sub_{r, s}$ includes precise start and end timestamps $(\tau^{\text{start}}_{r, s}, \tau^{\text{end}}_{r, s})$ that delineate the temporal extent of the associated motion description $\texttt{motion}_{r,s}$.
We use CLIP~\cite{radford2021learning} text encoder to first extract the textual features $emb_{\text{r}}$ for each motion subtitle $sub_{r, s}$:
\begin{equation}
emb_{\text{r}} = \texttt{Linear}(\texttt{CLIP}(\texttt{motion}_{r, s})) \in \mathbb{R}^D,
\end{equation}
where $\texttt{Linear}(\cdot)$ denotes a linear projection, $D$ denotes the dimension of the latent space.
To align these embeddings with the temporal structure of the animation sequence, each $emb{r, s}$ is replicated across its designated time interval $[\tau^{\text{start}}{r, s}, \tau^{\text{end}}{r, s}]$.
A learnable relative positional encoding is added to incorporate temporal awareness:
\begin{equation}
f_{r, i} = \begin{cases}
    emb_{\text{r, s}} + \texttt{PE}(i - \tau^{\text{start}}_{r, s}), & \text{if } i \in [\tau^{\text{start}}_{r, s}, \tau^{\text{end}}_{r, s}] \\
    \mathbf{0} & \text{otherwise}
\end{cases},
\end{equation}
where $s \in \{1, \dots, n\}$, and $\texttt{PE}(\cdot):\mathbb{Z}\rightarrow\mathbb{R}^D$ denotes a learnable function that encodes the relative temporal offset within the subtitle window.
These time-aligned region-specific semantic features $f_r \in \mathbb{R}^{L \times D}$ are then fed into the corresponding mapper networks to generate region-specific motion intensities or pose:
\begin{equation}
\begin{aligned}
\hat{\theta}^{\text{neck}} &= \texttt{Mapper}_{\text{neck}}(\mathbf{z}, f_{\text{neck}}), \\
\hat{I}_{\mathbf{L}_{\text{r}}} &= \texttt{Mapper}_{\text{r}}(\mathbf{z}, f_{\text{r}}), r \in \{\text{eyebrows}, \text{eyes}, \text{mouth}\}.
\end{aligned}
\end{equation}
This hierarchical architecture balances global coherence and localized control.
The shared audio-driven feature ensures consistent prosodic grounding across facial components, while the fine-grained motion subtitles provide region-specific semantic guidance.
Together, this design enables the synthesis of expressive, temporally aligned, and controllable singing-driven facial animation.

\begin{figure}[t]
    \centering
    \includegraphics[width=.5\textwidth]{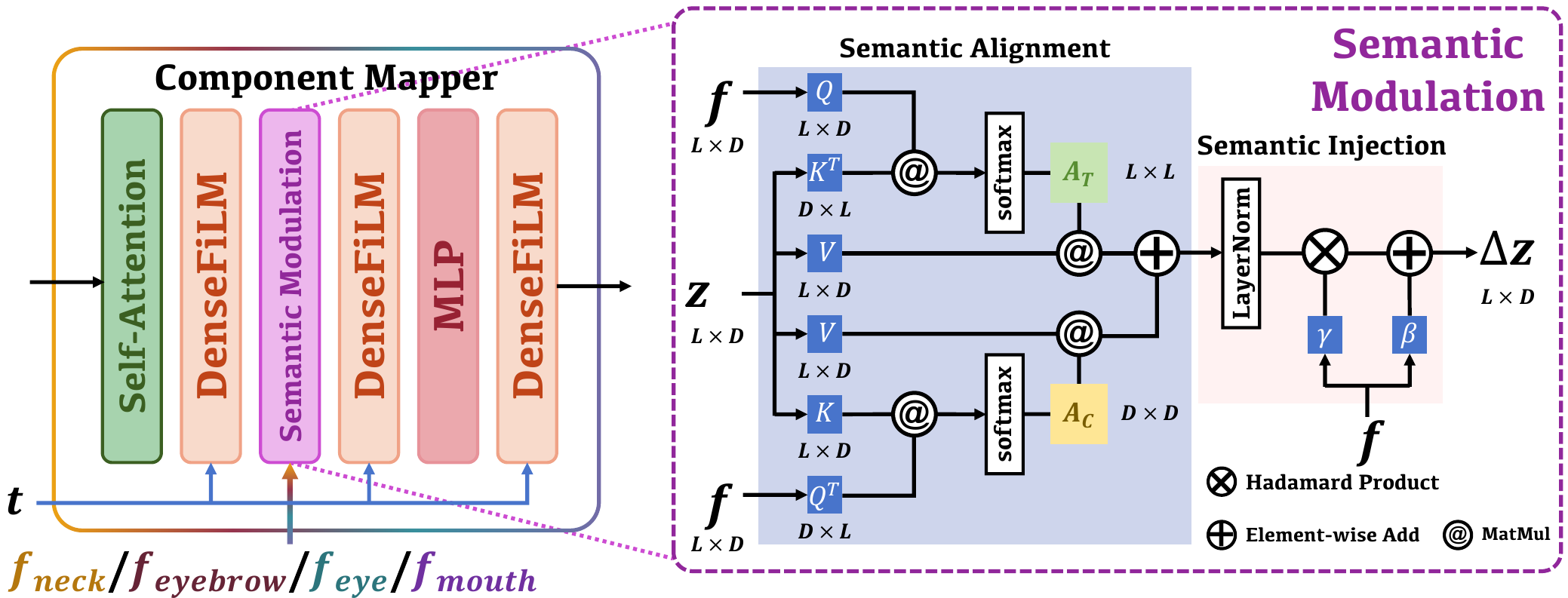}
    \caption{
        \textbf{Illustration of the Component Mapper.} 
        This module integrates region-wise motion subtitle features into the latent sequence via a Semantic Modulation layer.
        The layer consists two components: Semantic Alignment, which employs dual attention along temporal and channel dimensions to achieve precise feature alignment, and Semantic Injection, which modulates the latent representation through an AdaLN~\cite{peebles2023scalable} operation.
        } 
    \vspace{-10pt}
    \label{component_mapper}	
  \end{figure}
To enable fine-grained control over facial motion through the integration of region-specific semantic features, we devise a Semantic Modulation layer in the mapper network, as illustrated in~\reffig{component_mapper}.
The Semantic Modulation layer integrates region-specific semantic information into the latent sequence by first aligning the latent sequence with region-specific semantics through Semantic Alignment, and then enhancing facial motion control and accuracy by incorporating these aligned features via Semantic Injection, ensuring both temporal and semantic coherence in the resulting motion.
In the Semantic Alignment, we compute dual attention across both the temporal and channel dimensions to align the input latent sequence $\mathbf{z} \in \mathbb{R}^{L \times D}$ with the region-specific semantic features $f_r$.
In temporal attention, we only attend to the time interval where the motion subtitles are defined, ensuring that the attention mechanism focuses on relevant temporal segments.
The temporal attention mask $M_r$ is defined as follows:
\begin{equation}
M_r[i, j] =
\begin{cases}
0, \text{if } f_{r, i} \neq 0 \text{ and } f_{r, j} \neq 0 \\
-\infty, \text{otherwise}
\end{cases}, \forall i, j \in \{1, \dots, L\}
\end{equation}
The temporal attention matrix $A_T$ is computed as:
\begin{equation}
\begin{aligned}
Q_T &= f_r W_{T,Q} \in \mathbb{R}^{L \times D}, \quad K_T = \mathbf{z} W_{T,K} \in \mathbb{R}^{L \times D}, \\
A_T &= \texttt{softmax}\left( \frac{Q_T {K_T}^\top + M_r}{\sqrt{D}} \right) \in \mathbb{R}^{L \times L},
\end{aligned}
\end{equation}
The channel attention matrix $A_C$ is computed as:
\begin{equation}
\begin{aligned}
Q_C &= f_r W_{C,Q} \in \mathbb{R}^{L \times D}, \quad K_C = \mathbf{z} W_{C,K} \in \mathbb{R}^{L \times D}, \\
A_C &= \texttt{softmax}\left( \frac{{Q_C}^\top K_C}{\sqrt{L}} \right) \in \mathbb{R}^{D \times D}.
\end{aligned}
\end{equation}
The aligned semantic representation is then obtained via weighted aggregation using both attention maps:
\begin{equation}
\begin{aligned}
V_T &= \mathbf{z} W_{V,T}, \quad V_C = \mathbf{z} W_{V,C}, \\
\tilde{f} &= A_T V_z + A_C V_f.
\end{aligned}
\end{equation}
Next, in the Semantic Injection, the input $\mathbf{z}$ is modulated by the aligned features $\tilde{f}$ via an AdaLN~\cite{peebles2023scalable} operation. First, we apply layer normalization to $\tilde{f}$, and then perform an affine transformation where the scale $\gamma$ and shift $\beta$ parameters are learned from the semantic features $f_r$:
\begin{equation}
\begin{aligned}
\Delta\mathbf{z} &= \gamma(f_r) \otimes \texttt{LayerNorm}(\tilde{f}) + \beta(f_r), \\
\mathbf{z}_{\text{out}} &= \mathbf{z} + \Delta\mathbf{z}
\end{aligned}
\end{equation}
where $\otimes$ denotes Hadamard product.
This design enables the mapper network to dynamically inject fine-grained regional semantics into the latent sequence, facilitating temporally and semantically coherent facial motion generation.

Finally, the predicted motion intensities are converted into FLAME parameters via the Intensity2Motion predictor, as detailed in \refeq{eq:intensity2motion}:
\begin{equation}
[\hat{\psi}_{1:L}, \hat{\theta}^{\text{jaw}}_{1:L}] = f_\phi([\hat{I}_{\mathbf{L}_{\text{eyebrows}}}, \hat{I}_{\mathbf{L}_{\text{eyes}}}, \hat{I}_{\mathbf{L}_{\text{mouth}}}]).
\end{equation}

\subsection{Loss Functions}
We employ three loss terms to govern the training of our framework: an L2 reconstruction loss, a velocity loss, and an acceleration loss,  all applied to the outputs of both the mapper and the Intensity2Motion predictor:
\begin{equation}
\begin{aligned}
\mathcal{L}_{\text{recon}} &= \left\lVert \hat{y} - y \right\rVert_2^2, \mathcal{L}_{\text{vel}} = \left\lVert \hat{y}_{t+1} - \hat{y}_t \right\rVert_2^2, \\
\mathcal{L}_{\text{acc}} &= \left\lVert \hat{y}_{t+2} - 2\hat{y}_{t+1} + \hat{y}_t \right\rVert_2^2,
\end{aligned}
\end{equation}
where $\hat{y}$ and $y$ denote the prediction and ground truth.
To balance the contributions of these loss terms, we use a gradient-guided adaptive loss weighting mechanism.
The total loss is computed as a weighted sum of the individual loss terms:
\begin{equation}
\mathcal{L}_{\text{total}} = \lambda_{\text{recon}} \cdot \mathcal{L}_{\text{recon}} + \lambda_{\text{vel}} \cdot \mathcal{L}_{\text{vel}} + \lambda_{\text{acc}} \cdot \mathcal{L}_{\text{acc}},
\end{equation}
where the weights $\lambda$ are adaptively computed based on the gradient magnitudes to balance the contributions of each term.
During training, annotated motion subtitles are used to supervise the mapper networks, ensuring that the generated motion intensities align with the expected facial dynamics.
For inference, motion subtitles generated by Sing-CoT and AGRA are utilized to guide the generation.

\vspace{-10pt}
\section{Experiments}
\begin{table*}[t]
    \label{tab:comparison}
    \centering
    \caption{\textbf{Quantitative comparisons with state-of-the-art methods.}
    \textbf{Bold} indicates the best result, and \underline{underline} indicates the second best.
    $\downarrow$ means lower is better, $\uparrow$ means higher is better, and $\rightarrow$ means closer to the GT is better.
    LVE, FVE are reported in $mm$, and FDD in $1 \times 10^{-2} mm$.
    Note that artificially low LVE and FVE scores may result from unnatural motion sequences with excessively high freeze rates.
    $\text{FID}_{fm}$, $\text{FID}_{\Delta fm}$ and SND are applicable for FLAME parameterized methods, while BA are not utlized for methods without neck pose modules.
    The results demonstrate the superiority of our approach.
    }  
        \begin{tabular}{c|ccc|ccc|c|c|c} \hline
        Methods                                                 & LVE $\downarrow$    & FVE $\downarrow$    & Freeze Rate $\downarrow$    & $\text{FID}_{fm} \downarrow$    & $\text{FID}_{\Delta fm} \downarrow$    & SND $\downarrow$    & FDD $\rightarrow$      & BA $\rightarrow$   & Train Params $\downarrow$  \\ \hline
        GT                                                      & -                   & -                   & -                           & -                               & -                                      & -                   & 0.0000                 & 0.2379             & -                          \\
        mean GT                                                 & 5.4617              & 1.2473              & 100\%                       & -                               & -                                      & -                   & -                      & -                  & -                          \\ \hline
        FaceFormer~\cite{fan2022faceformer}                     & 8.3815              & 2.2256              & 51.6689\%                   & -                               & -                                      & -                   & 3.6631                 & -                  & 92.22 M                    \\
        CodeTalker~\cite{xing2023codetalker}                    & 9.1110              & 2.2654              & 11.5437\%                   & -                               & -                                      & -                   & 3.4876                 & -                  & 314.70 M                   \\
        Imitator~\cite{thambiraja2023imitator}                  & \textbf{7.1733}     & \textbf{1.8205}     & 67.7134\%                   & -                               & -                                      & -                   & 3.8349                 & -                  & 91.27 M                    \\
        SelfTalk~\cite{peng2023selftalk}                        & 8.8820              & 2.4673              & 21.0643\%                   & -                               & -                                      & -                   & 2.5038                 & -                  & 449.46 M                   \\ 
        LG-LDM~\cite{song2024expressive}                        & 9.4266              & 2.2676              & 5.6680\%                    & -                               & -                                      & -                   & 4.3390                 & -                  & 471.52 M                   \\ \hline
        FaceFormer~(FLAME ver.)~\cite{fan2022faceformer}        & 9.6633              & 4.6802              & 47.0000\%                   & 51.6832                         & 0.8716                                 & 52.5548             & 4.6802                 & 0.2218             & 90.28 M                    \\
        CodeTalker~(FLAME ver.)~\cite{xing2023codetalker}       & 10.7223             & 2.7418              & 22.7277\%                   & 69.7111                         & 0.5785                                 & 70.2896             & -3.6581                & 0.1521             & 268.57 M                   \\
        Imitator~(FLAME ver.)~\cite{thambiraja2023imitator}     & 9.4697              & 2.9034              & 70.8456\%                   & 89.3647                         & 0.8907                                 & 90.2554             & 3.7435                 & \underline{0.2252} & 90.29 M                    \\
        SelfTalk~(FLAME ver.)~\cite{peng2023selftalk}           & 10.2898             & 2.6210              & 57.8557\%                   & 49.1636                         & 0.7019                                 & 49.8656             & 4.1252                 & 0.2505             & 441.76 M                   \\ 
        FaceDiffuser~\cite{stan2023facediffuser}                & 11.5882             & 2.8015              & \underline{1.6593\%}        & 30.9131                         & \underline{0.3360}                     & 31.2492             & \underline{0.9744}     & 0.2509             & 81.09 M                    \\ 
        LG-LDM~(FLAME ver.)~\cite{song2024expressive}           & 19.7557             & 12.6928             & 15.2721\%                   & \underline{20.8294}             & 0.3590                                 & \underline{21.1884} & 36.9100                & 0.2539             & 440.77 M                   \\ 
        DEEPTalk~\cite{kim2025deeptalk}                         & 11.5723             & 2.8151              & \textbf{0.0030\%}           & 47.6321                         & 1.9150                                 & 49.5471             & 1.5381                 & 0.2162             & \textbf{8.52 M}            \\ \hline
        \rowcolor{gray!20}
        Ours                                                    & \underline{8.3036}  & \underline{2.1364}  & 8.6919\%                    & \textbf{4.8187}                 & \textbf{0.0671}                        & \textbf{4.8858}     & \textbf{0.1611}        & \textbf{0.2472}    & \underline{22.14 M}        \\ \hline
        \end{tabular}
    \label{comparison}
\end{table*}
\subsection{Settings}
\subsubsection{Implementation Details}
We conduct our experiments using the collected dataset SingMoSub.
The dataset is split into training and testing sets with a 9:1 ratio.
As certain segments within the same song may be repeated, we make sure that the songs in the training and testing sets do not overlap to prevent any potential data leakage.
We use Gemini 2.5 Flash~\cite{comanici2025gemini} as the LLM for generating motion subtitles.
For training, we use the Adam~\cite{xie2024adan} optimizer with a learning rate of $4 \times 10^{-4}$ and a batch size of 64.
The model is trained with 4 NVIDIA RTX 4090 GPUs.

\begin{table*}[t]
    \label{tab:ablation_study}
    \centering
    \caption{
    \textbf{Ablation studies on motion subtitles \& intensity proxy.}
    \textbf{Bold} indicates the best result, and \underline{underline} indicates the second best.
    $\downarrow$ means lower is better, $\uparrow$ means higher is better, and $\rightarrow$ means closer to the ground truth is better.
    LVE, FVE are measured in $mm$, and FDD in $1 \times 10^{-2} mm$.
    The results confirm the effectiveness of the proposed motion subtitles, intensity proxy, AGRA and Sing-CoT.
    }  
        \begin{tabular}{c|ccc|ccc|c|c} \hline
        Methods               & LVE $\downarrow$    & FVE $\downarrow$    & Freeze Rate $\downarrow$    & $\text{FID}_{fm} \downarrow$    & $\text{FID}_{\Delta fm} \downarrow$    & SND $\downarrow$    & FDD $\rightarrow$      & BA $\rightarrow$   \\ \hline
        GT Sub                & \textbf{7.8513}     & \textbf{1.9537}     & 9.8180 \%                   & \textbf{4.7649}                 & \underline{0.0818}                     & \textbf{4.8467}     & \underline{0.3508}     & \underline{0.2462} \\
        Lyrics Sub            & 9.0268              & 2.3345              & 18.2900 \%                  & 8.6098                          & 0.2354                                 & 8.8452              & 0.4740                 & 0.2150             \\ 
        Direct FLAME          & 9.4185              & 2.3548              & 14.3094 \%                  & 12.1793                         & 0.2732                                 & 12.4525             & 1.6416                 & \textbf{0.2367}    \\ 
        Direct Vert           & 9.3905              & 10.2193             & 13.0572 \%                  & -                               & -                                      & -                   & 4.2433                 & -                  \\ \hline
        \rowcolor{gray!20}
        Ours                  & \underline{8.3036}  & \underline{2.1364}  & \textbf{8.6919 \%}          & \underline{4.8187}              & \textbf{0.0671}                        & \underline{4.8858}  & \textbf{0.1611}        & 0.2472             \\ \hline
        \end{tabular}
    \label{ablation_study}
\end{table*}
\subsubsection{Metrics}
We evaluate the generated results from four aspects: 

(1) \textit{Lip synchronization \& facial accuracy}:
Following existing works~\cite{fan2022faceformer,xing2023codetalker,thambiraja2023imitator,peng2023selftalk,song2024expressive,stan2023facediffuser,kim2025deeptalk}, we assess lip synchronization and overall facial geometry accuracy by measuring the difference between the generated and ground truth vertices.
To evaluate lip synchronization, we use the \textbf{Lip Vertex Error~(LVE)}, which calculates the average maximum L2 loss of lip vertices in each frame, quantifying the accuracy of lip synchronization.
The \textbf{Face Vertex Error~(FVE)} measures the average L2 loss of face region vertices across the entire sequence, evaluating the overall facial alignment and shape.
However, some methods generate static outputs that regress toward the average pose, resulting in \textbf{artificially low LVE and FVE scores} that fail to capture dynamic facial expressions.
As shown in~\reftab{comparison}, using a static sequence based on the average ground truth frame resulted in the lowest LVE and FVE, confirming this issue.
To address this, we propose the \textbf{Freeze Rate}, which evaluates the percentage of frames below the minimal movement, indicating frozen, static animations.
The threshold for minimal movement is set to the mean vertex displacement across each ground truth sequence.
Higher Freeze Rate values suggest a failure to capture dynamic facial expressions, thereby deteriorating lip synchronization and facial accuracy.

(2) \textit{Temporal Coherence and Quality}:
The \textbf{$\text{FID}_{fm}$} metric computes the Fréchet Inception Distance of the FLAME parameters, assessing the similarity between the generated and ground truth FLAME parameter distributions.
Additionally, \textbf{$\text{FID}_{\Delta fm}$} evaluates the temporal coherence by measuring the FID of the difference in FLAME parameters between consecutive frames, with a lower score indicating smoother transitions. 
Finally, the \textbf{SND} is the sum of $\text{FID}_{fm}$ and $\text{FID}_{\Delta fm}$, providing a combined measure of both the overall quality and temporal smoothness of the generated facial motion. 
Together, these metrics offer a comprehensive evaluation of both the spatial and temporal aspects of the generated animation, ensuring that the facial motion is not only accurate but also consistent and smooth across the sequence.

(3) \textit{Emotional expressiveness}:
Following CodeTalker~\cite{xing2023codetalker}, we measure facial dynamics variation using the \textbf{Upper-face Dynamics Deviation~(FDD)}, which captures the expressive movement in the upper face essential for emotional conveyance.

(4) \textit{Music alignment}:
Unlike talking head generation, where the focus is on lip-synchronizing with speech, singing-driven head animation must also align with the rhythm and beat of the music.
To evaluate the alignment with music, we adopt the \textbf{Beat Alignment~(BA) score} frequently used from music-to-dance generation~\cite{huang2024beat,siyao2022bailando,tseng2023edge}.
This metric measures the alignment of facial landmarks with the music's beats, ensuring rhythmic coherence and enhancing the naturalness of the performance.

\subsection{Comparison with Existing Methods}
Since very few existing works have investigated the 3D singing head animation problem, we incorporate several state-of-the-art, open-source audio-driven talking head generation methods for comparison, including both vertex-based~\cite{fan2022faceformer, xing2023codetalker,thambiraja2023imitator, peng2023selftalk} and FLAME-based~\cite{stan2023facediffuser,kim2025deeptalk} approaches.
Our experiments show that vertex-based methods, due to the large variation in neck poses in our dataset, result in generated outputs that regress toward the average, causing them to freeze and appear static.
To address this issue, we set the neck pose to zero during training, focusing solely on facial animation without head movement.
For fairness, we also train a FLAME-parameterized variant of these approaches, where the output layer is modified to predict FLAME parameters, including the neck pose, instead of vertex offsets.
All methods are trained on the SingMoSub to ensure a fair comparison.
This work focuses exclusively on the generation of 3D head motion.
To eliminate shape-related variations caused by individual identity, all experiments are conducted using a fixed dummy shape parameter set to all zeros.
\subsubsection{Quantitative Comparisons}
We first compare our method with the state-of-the-art approaches quantitatively.
The results, summarized in~\reftab{comparison}, demonstrate that our method consistently outperforms existing approaches in lip synchronization, facial accuracy, emotional expressiveness, and music alignment.
By strategically leveraging LLM priors, our model achieves optimal performance while maintaining a considerable compact model size.
Notably, while Imitator~\cite{thambiraja2023imitator} achieves the lowest LVE and FVE, it exhibits the \textbf{highest Freeze Rate}, indicating that the generated facial animations are static and lack dynamic expression. This suggests that Imitator struggles to capture the subtle variations in facial expressions necessary for realistic emotional conveyance.
In contrast, DEEPTalk~\cite{kim2025deeptalk} achieves the lowest Freeze Rate, indicating larger dynamic range, but its \textbf{high FVE and LVE} indicate severe issues with facial geometry, as well as jittering artifacts that compromise the realism of the generated facial movements.
Our method \textbf{strikes the optimal balance} between LVE, FVE, and Freeze Rate, showcasing our ability to generate head animations that are not only dynamic but also maintain accurate lip synchronization and facial geometry.
This balance ensures that our generated expressions are both realistic and emotionally expressive, surpassing the limitations of previous methods that either sacrifice expressiveness for accuracy or vice versa.
Regarding temporal coherence and overall quality, our method significantly outperforms the other competitives, with an SND that is 76.9\% lower than the second-best method~(4.8858 vs. 21.1884).
This demonstrates that our approach not only ensures the overall accuracy of the predictions but also maintains smooth transitions across frames, preserving temporal consistency and preventing abrupt or unnatural changes in movements.
For emotional expressiveness, our method leads with the best FDD score, outperforming the second-best by a significant margin~(0.1611 vs. 0.9744).
This highlights our method's capacity to generate emotionally rich upper-face movements, essential for conveying emotions.
In terms of music alignment, we achieve the best BA score, confirming that our method effectively synchronizes head motion with the music rhythm, a critical aspect for singing scenario.

\begin{figure}[t]
    \centering
    \includegraphics[width=\linewidth]{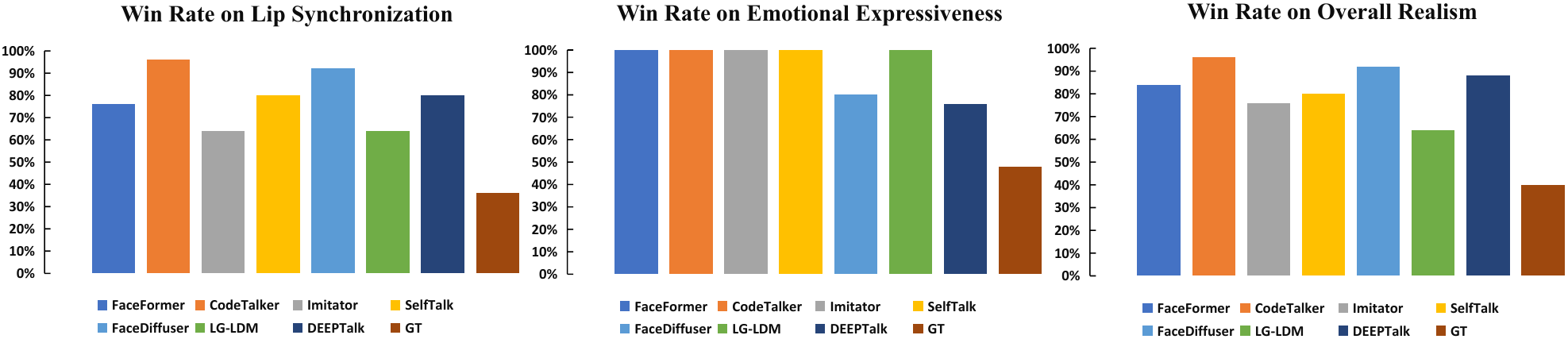}
    \caption{
        \textbf{User Study.} Our method achieves the highest preference in terms of lip synchronization, emotional expressiveness, and overall realism compared to other methods.
        }
    \vspace{-10pt}
    \label{user_study}	
  \end{figure} 
\begin{figure*}[t]
\centering
\includegraphics[width=\linewidth,scale=1.00]{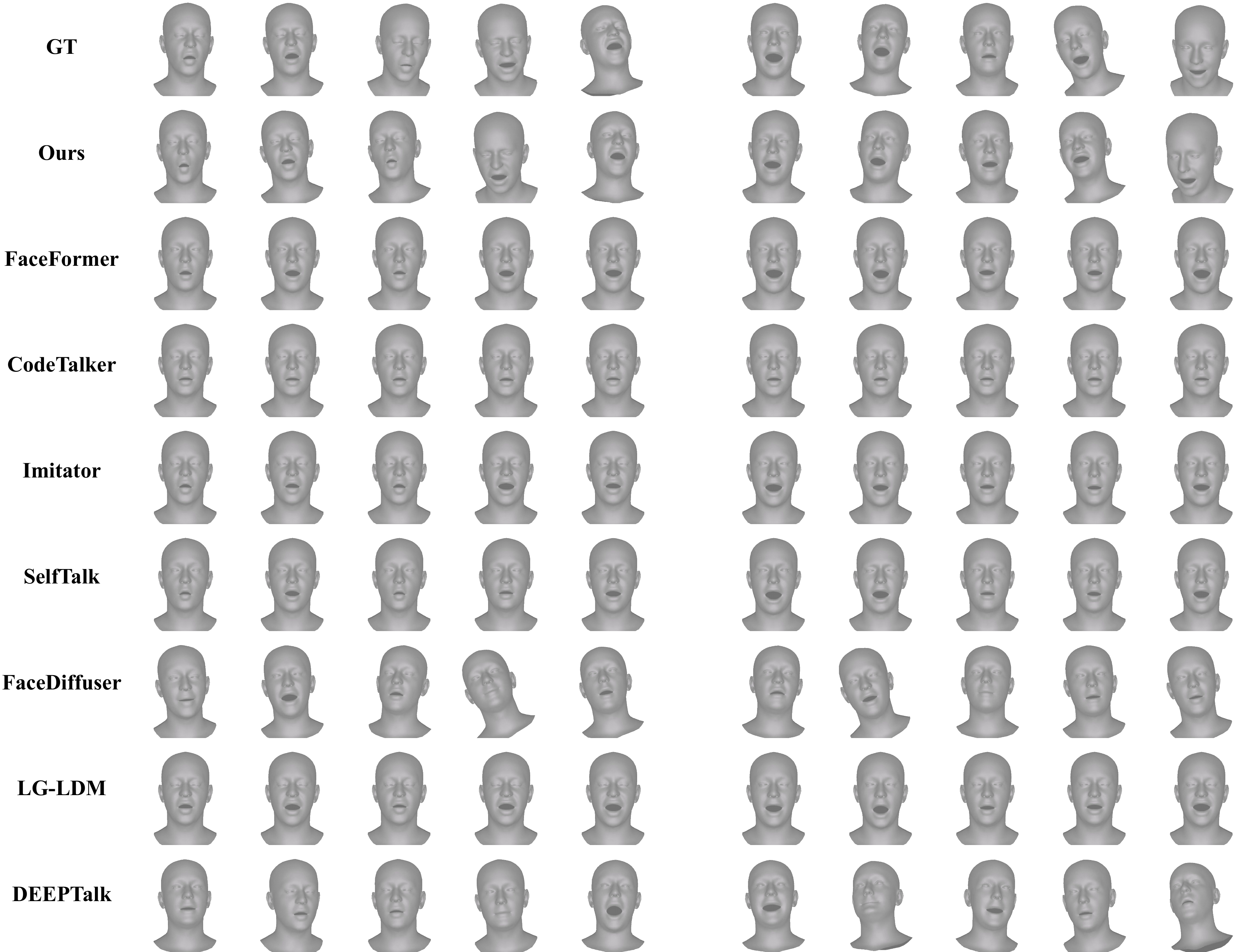}
\caption{
    \textbf{Qualitative comparison with other methods.}
    The results demonstrate the superiority of our approach in generating realistic and emotional expressive head animations.
} 
\vspace{-10pt}
\label{qualitative_comparison}	
\end{figure*}

\begin{figure}[t]
    \centering
    \includegraphics[width=.40\textwidth]{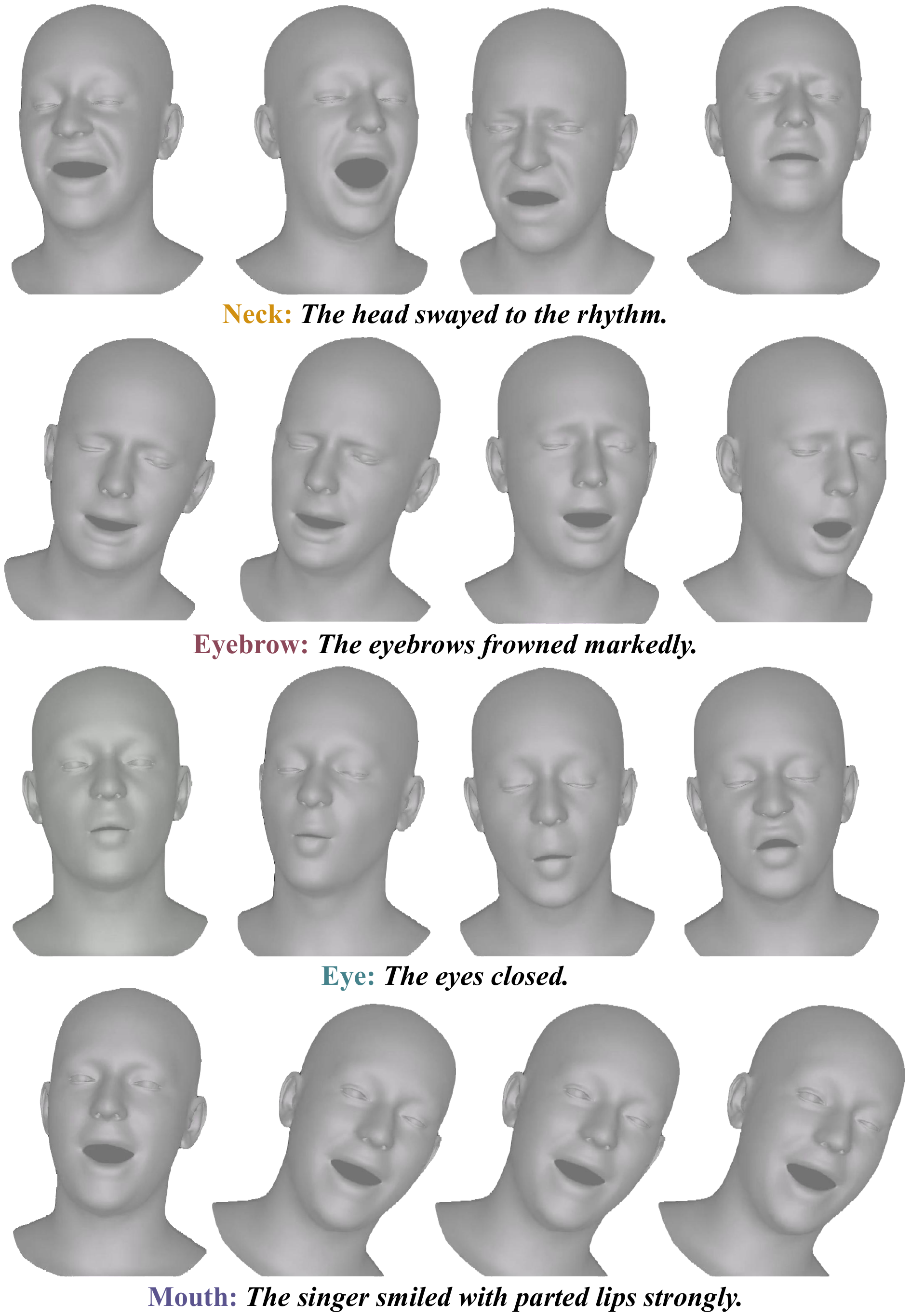}
    \caption{
        \textbf{Visualization for region-wise motion subtitles.}
        Our method enables fine-grained control over head animation by leveraging motion subtitles corresponding to specific regions, resulting in more expressive and emotionally rich 3D head motions.
        } 
    \label{regionwise_sub}	
  \end{figure}
\subsubsection{Qualitative Comparisons}
We also conduct qualitative comparisons with existing methods.
Visualized results in~\reffig{qualitative_comparison} demonstrate that our method consistently generates more realistic and expressive facial animations than competing approaches.
We further visualize the generated motions alongside the corresponding region-wise motion subtitles in \reffig{regionwise_sub}.
For clarity, each region is presented with its textual description and the motion it governs during the relevant temporal segment.

For a more comprehensive comparison, we incorporate a user study with 25 participants to evaluate the 35 random sampled generated results based on the lip synchronization, emotional expressiveness, and overall realism.
The results, shown in~\reffig{user_study}, indicate that our method is preferred over the other methods in all three aspects.
Specifically, we found that participants rated our method significantly higher in terms of expressiveness and overall realism compared to the baseline methods.
Additional visualizations are available on the \href{https://zikaihuangscut.github.io/Think2Sing/}{\textcolor{blue}{project page}}.

\subsection{Ablation Studies}
In this section, we present ablation studies to evaluate the contributions of key components in our method, including motion subtitles, intensity proxy, and the integration of AGRA and Sing-CoT.

\textbf{Effectiveness of Motion Subtitles.}
We perform ablation studies to assess the effectiveness of motion subtitles in our method.
Specifically, we compare our approach with two variants: one using only lyrics text as the condition~(denoted as Lyrics Sub) and the other using annotated motion subtitles~(denoted as GT Sub) as the condition.
As shown in~\reftab{ablation_study}, the use of motion subtitles significantly improves performance across all evaluation metrics compared to using lyrics text alone.
Lyrics Sub. exhibits significant degradation in all key metrics.
This performance drop is likely due to the lack of clear guidance provided by the lyrics text, which fails to capture the nuanced facial motions required for expressive animation. In contrast, motion subtitles offer more precise and actionable cues, allowing the model to more effectively capture and generate expressive facial movements.
Furthermore, GT Sub results in even better lip synchronization and facial accuracy.
However, our method, which combines AGRA and Sing-CoT, achieves performance comparable to that of the GT Sub variant across all metrics.
This demonstrates the effectiveness of our proposed framework in approximating ground-truth-level performance, validating the utility of both AGRA and Sing-CoT in generating expressive and temporally coherent motion subtitles.

\textbf{Effectiveness of Intensity Proxy.}
We evaluate the effectiveness of using intensity as a proxy by training a vertex-based variant~(denoted as Direct Vert) and a FLAME-based variant~(denoted as Direct FLAME).
To ensure a fair comparison, we match the model size of each variant with the proposed method.
As shown in~\reftab{ablation_study}, the Direct FLAME variant performs poorly across all evaluation metrics, particularly in terms of $\text{FID}_{fm}$.
This performance degradation can be attributed to the highly coupled nature of the FLAME parameters.
The vertex-based variant shows  considerable degradations in all metrics as well, showcasing the effectiveness of using intensity as a proxy.
Without the intensity proxy, the model struggles to generate region-specific facial motion, as the motion subtitles cannot be effectively routed to control specific regions.
This results in increased ambiguity and difficulty in achieving fine-grained control over facial dynamics.
In contrast, using intensity as a proxy allows for more precise and targeted control, leading to better performance across all metrics.

\begin{table}[t]
  
\begin{minipage}{\linewidth}
    \centering
    \captionof{table}{
        \textbf{Ablation studies on AGRA \& Sing-CoT in terms of the number of rounds LLM takes for successful validations.}
        \textbf{Bold} indicates best result. $\uparrow$ means higher is better.
    }
        {
            \renewcommand\arraystretch{1.1}
            \begin{tabular}{c|ccc} \hline
                Methods                   & 1\textsuperscript{st} Pass $\uparrow$    & 2\textsuperscript{nd} Pass $\uparrow$    & 3\textsuperscript{rd} Pass $\uparrow$ \\ \hline
                Baseline                  & 14.18 \%                                 & -                                        & -                                     \\
                Lyrics RAG Only           & 18.85 \%                                 & -                                        & -                                     \\
                AGRA Only                 & 22.95 \%                                 & -                                        & -                                     \\
                Sing-CoT Only             & 73.13 \%                                 & 90.30 \%                                 & 95.52\%                               \\ \hline
                \rowcolor{gray!20}
                AGRA + Sing-CoT~(Ours)    & \textbf{78.35 \%}                        & \textbf{91.04 \%}                        & \textbf{96.27 \%}                     \\ \hline
            \end{tabular}
            \label{rag_cot_ablation}
    }
\end{minipage} 

\end{table}
\textbf{Effectiveness of AGRA and Sing-CoT.}
We further investigate the individual contributions of AGRA and Sing-CoT by comparing our full model with four variants:
1) baseline model that does not use AGRA or Sing-CoT~(baseline);
2) model using only naive lyrics RAG~(Lyrics RAG only);
3) model using only AGRA~(AGRA only);
4) model using only Sing-CoT~(Sing-CoT only).
We evaluate their performance based on the number of rounds LLM takes for successful validations.
As shown in~\reftab{rag_cot_ablation}, compared to the baseline, incorporating Sing-CoT dramatically enhances the model's performance, increasing the validation success rate in the first round from 14.18\% to 73.13\%.
Compared with using naive lyrics RAG, using AGRA further boosts the performance, raising the first-round success rate to 22.95\%.
When both AGRA and Sing-CoT are combined, the LLM benefits from the strengths of both components, achieving the highest success rates in each round.
The third-round success rate reaches 96.27\%, effectively improving the inference capability of the LLM.
\section{Limitation}
Similar to other methods~\cite{wu2023singinghead,wu2024mmhead} built on the FLAME mesh, our approach cannot model phenomena such as eye-gaze dynamics or hair motion, due to the inherent limitations of the FLAME representation. Another limitation is the latency introduced by LLM processing during motion subtitle generation. However, once the subtitles are generated offline, our method can achieve an efficient inference speed of over 200 FPS on a single NVIDIA RTX 4090.

\section{Conclusion}
In this work, we presented Think2Sing, the first LLM-assisted framework for expressive singing-driven 3D head animation. Unlike prior methods that rely on direct audio-to-motion mappings, our approach leverages structured motion subtitles, inferred through an LLM-assisted Sing-CoT reasoning scheme with acoustic-guided retrieval, to provide explicit, fine-grained, and temporally aligned motion guidance. We further introduced a motion intensity representation that serves as a proxy between audio and FLAME parameters, effectively reducing learning ambiguity, enabling region-wise control, and improving the modeling of subtle facial dynamics. Moreover, we curated SingMoSub, the first multimodal singing dataset with synchronized clips, acoustic descriptors, and region-wise motion subtitles, offering rich supervision for semantic and prosodic learning. Extensive quantitative and qualitative experiments demonstrate that Think2Sing surpasses state-of-the-art methods in producing realistic and expressive motions, while enabling flexible subtitle-conditioned editing for precise and user-controllable animation synthesis.


%





\ifCLASSOPTIONcaptionsoff
  \newpage
\fi

\bibliographystyle{IEEEtran}
\bibliography{main}

\begin{thebibliography}{10}
\providecommand{\url}[1]{#1}
\csname url@samestyle\endcsname
\providecommand{\newblock}{\relax}
\providecommand{\bibinfo}[2]{#2}
\providecommand{\BIBentrySTDinterwordspacing}{\spaceskip=0pt\relax}
\providecommand{\BIBentryALTinterwordstretchfactor}{4}
\providecommand{\BIBentryALTinterwordspacing}{\spaceskip=\fontdimen2\font plus
\BIBentryALTinterwordstretchfactor\fontdimen3\font minus \fontdimen4\font\relax}
\providecommand{\BIBforeignlanguage}[2]{{%
\expandafter\ifx\csname l@#1\endcsname\relax
\typeout{** WARNING: IEEEtran.bst: No hyphenation pattern has been}%
\typeout{** loaded for the language `#1'. Using the pattern for}%
\typeout{** the default language instead.}%
\else
\language=\csname l@#1\endcsname
\fi
#2}}
\providecommand{\BIBdecl}{\relax}
\BIBdecl

\bibitem{yu20193d}
J.~Yu, C.~W. Chen, and Z.~Wang, ``3d singing head for music vr: Learning external and internal articulatory synchronicity from lyric, audio and notes,'' in \emph{Proceedings of the 27th ACM International Conference on Multimedia}, 2019, pp. 945--952.

\bibitem{wu2023singinghead}
S.~Wu, Y.~Li, W.~Zhang, J.~Jia, Y.~Zhu, Y.~Yan, G.~Zhai, and X.~Yang, ``Singinghead: A large-scale 4d dataset for singing head animation,'' \emph{arXiv preprint arXiv:2312.04369}, 2023.

\bibitem{liu2024musicface}
P.~Liu, W.~Deng, H.~Li, J.~Wang, Y.~Zheng, Y.~Ding, X.~Guo, and M.~Zeng, ``Musicface: Music-driven expressive singing face synthesis,'' \emph{Computational Visual Media}, vol.~10, no.~1, pp. 119--136, 2024.

\bibitem{xie2025let}
X.~Xie, Z.~Huang, W.~Xu, P.~Xiao, X.~Xu, and H.~Zhang, ``Let's chorus: Partner-aware hybrid song-driven 3d head animation,'' in \emph{Proceedings of the Computer Vision and Pattern Recognition Conference}, 2025, pp. 5467--5476.

\bibitem{quinto2013emotional}
L.~Quinto, W.~F. Thompson, and F.~L. Keating, ``Emotional communication in speech and music: The role of melodic and rhythmic contrasts,'' \emph{Frontiers in psychology}, vol.~4, p. 184, 2013.

\bibitem{livingstone2013acoustic}
S.~R. Livingstone, K.~Peck, and F.~A. Russo, ``Acoustic differences in the speaking and singing voice,'' in \emph{Proceedings of Meetings on Acoustics}, vol.~19, no.~1.\hskip 1em plus 0.5em minus 0.4em\relax Acoustical Society of America, 2013, p. 035080.

\bibitem{eyben2015emotion}
F.~Eyben, G.~L. Salom{\~a}o, J.~Sundberg, K.~R. Scherer, and B.~W. Schuller, ``Emotion in the singing voice—a deeperlook at acoustic features in the light ofautomatic classification,'' \emph{EURASIP Journal on Audio, Speech, and Music Processing}, vol. 2015, no.~1, p.~19, 2015.

\bibitem{livingstone2015common}
S.~R. Livingstone, W.~F. Thompson, M.~M. Wanderley, and C.~Palmer, ``Common cues to emotion in the dynamic facial expressions of speech and song,'' \emph{Quarterly Journal of Experimental Psychology}, vol.~68, no.~5, pp. 952--970, 2015.

\bibitem{deng2025occlusion}
Y.~Deng, Y.~Lu, Y.~Xu, Y.~Nie, and S.~He, ``Occlusion-insensitive talking head video generation via facelet compensation,'' in \emph{Proceedings of the AAAI Conference on Artificial Intelligence}, vol.~39, no.~3, 2025, pp. 2726--2734.

\bibitem{li2024singer}
Y.~Li, Z.~Zhou, Z.~Wang, W.~Xue, W.~Luo, and Y.~Guo, ``Singer: Vivid audio-driven singing video generation with multi-scale spectral diffusion model,'' \emph{arXiv preprint arXiv:2412.03430}, 2024.

\bibitem{xie2024x}
Y.~Xie, H.~Xu, G.~Song, C.~Wang, Y.~Shi, and L.~Luo, ``X-portrait: Expressive portrait animation with hierarchical motion attention,'' in \emph{ACM SIGGRAPH 2024 Conference Papers}, 2024, pp. 1--11.

\bibitem{fan2022faceformer}
Y.~Fan, Z.~Lin, J.~Saito, W.~Wang, and T.~Komura, ``Faceformer: Speech-driven 3d facial animation with transformers,'' in \emph{Proceedings of the IEEE/CVF conference on computer vision and pattern recognition}, 2022, pp. 18\,770--18\,780.

\bibitem{xing2023codetalker}
J.~Xing, M.~Xia, Y.~Zhang, X.~Cun, J.~Wang, and T.-T. Wong, ``Codetalker: Speech-driven 3d facial animation with discrete motion prior,'' in \emph{Proceedings of the IEEE/CVF Conference on Computer Vision and Pattern Recognition}, 2023, pp. 12\,780--12\,790.

\bibitem{stan2023facediffuser}
S.~Stan, K.~I. Haque, and Z.~Yumak, ``Facediffuser: Speech-driven 3d facial animation synthesis using diffusion,'' in \emph{Proceedings of the 16th ACM SIGGRAPH Conference on Motion, Interaction and Games}, 2023, pp. 1--11.

\bibitem{peng2023selftalk}
Z.~Peng, Y.~Luo, Y.~Shi, H.~Xu, X.~Zhu, H.~Liu, J.~He, and Z.~Fan, ``Selftalk: A self-supervised commutative training diagram to comprehend 3d talking faces,'' in \emph{Proceedings of the 31st ACM International Conference on Multimedia}, 2023, pp. 5292--5301.

\bibitem{quinto2014singing}
L.~R. Quinto, W.~F. Thompson, C.~Kroos, and C.~Palmer, ``Singing emotionally: a study of pre-production, production, and post-production facial expressions,'' \emph{Frontiers in psychology}, vol.~5, p. 262, 2014.

\bibitem{edwards2016jali}
P.~Edwards, C.~Landreth, E.~Fiume, and K.~Singh, ``Jali: an animator-centric viseme model for expressive lip synchronization,'' \emph{ACM Transactions on graphics (TOG)}, vol.~35, no.~4, pp. 1--11, 2016.

\bibitem{taylor2017deep}
S.~Taylor, T.~Kim, Y.~Yue, M.~Mahler, J.~Krahe, A.~G. Rodriguez, J.~Hodgins, and I.~Matthews, ``A deep learning approach for generalized speech animation,'' \emph{ACM Transactions On Graphics (TOG)}, vol.~36, no.~4, pp. 1--11, 2017.

\bibitem{taylor2012dynamic}
S.~L. Taylor, M.~Mahler, B.-J. Theobald, and I.~Matthews, ``Dynamic units of visual speech,'' in \emph{Proceedings of the 11th ACM SIGGRAPH/Eurographics conference on Computer Animation}, 2012, pp. 275--284.

\bibitem{xu2013practical}
Y.~Xu, A.~W. Feng, S.~Marsella, and A.~Shapiro, ``A practical and configurable lip sync method for games,'' in \emph{Proceedings of Motion on Games}, 2013, pp. 131--140.

\bibitem{zhou2018visemenet}
Y.~Zhou, Z.~Xu, C.~Landreth, E.~Kalogerakis, S.~Maji, and K.~Singh, ``Visemenet: Audio-driven animator-centric speech animation,'' \emph{ACM Transactions on Graphics (ToG)}, vol.~37, no.~4, pp. 1--10, 2018.

\bibitem{vocaset}
D.~Cudeiro, T.~Bolkart, C.~Laidlaw, A.~Ranjan, and M.~J. Black, ``Capture, learning, and synthesis of 3d speaking styles,'' in \emph{Proceedings of the IEEE/CVF conference on computer vision and pattern recognition}, 2019, pp. 10\,101--10\,111.

\bibitem{wu2023speech}
H.~Wu, S.~Zhou, J.~Jia, J.~Xing, Q.~Wen, and X.~Wen, ``Speech-driven 3d face animation with composite and regional facial movements,'' in \emph{Proceedings of the 31st ACM International Conference on Multimedia}, 2023, pp. 6822--6830.

\bibitem{richard2021meshtalk}
A.~Richard, M.~Zollh{\"o}fer, Y.~Wen, F.~De~la Torre, and Y.~Sheikh, ``Meshtalk: 3d face animation from speech using cross-modality disentanglement,'' in \emph{Proceedings of the IEEE/CVF international conference on computer vision}, 2021, pp. 1173--1182.

\bibitem{karras2017audio}
T.~Karras, T.~Aila, S.~Laine, A.~Herva, and J.~Lehtinen, ``Audio-driven facial animation by joint end-to-end learning of pose and emotion,'' \emph{ACM Transactions on Graphics (ToG)}, vol.~36, no.~4, pp. 1--12, 2017.

\bibitem{danvevcek2023emotional}
R.~Dan{\v{e}}{\v{c}}ek, K.~Chhatre, S.~Tripathi, Y.~Wen, M.~Black, and T.~Bolkart, ``Emotional speech-driven animation with content-emotion disentanglement,'' in \emph{SIGGRAPH Asia 2023 Conference Papers}, 2023, pp. 1--13.

\bibitem{peng2023emotalk}
Z.~Peng, H.~Wu, Z.~Song, H.~Xu, X.~Zhu, J.~He, H.~Liu, and Z.~Fan, ``Emotalk: Speech-driven emotional disentanglement for 3d face animation,'' in \emph{Proceedings of the IEEE/CVF international conference on computer vision}, 2023, pp. 20\,687--20\,697.

\bibitem{wang2020mead}
K.~Wang, Q.~Wu, L.~Song, Z.~Yang, W.~Wu, C.~Qian, R.~He, Y.~Qiao, and C.~C. Loy, ``Mead: A large-scale audio-visual dataset for emotional talking-face generation,'' in \emph{European conference on computer vision}.\hskip 1em plus 0.5em minus 0.4em\relax Springer, 2020, pp. 700--717.

\bibitem{song2024expressive}
W.~Song, X.~Wang, Y.~Jiang, S.~Li, A.~Hao, X.~Hou, and H.~Qin, ``Expressive 3d facial animation generation based on local-to-global latent diffusion,'' \emph{IEEE Transactions on Visualization and Computer Graphics}, 2024.

\bibitem{liu2024emoface}
C.~Liu, Q.~Lin, Z.~Zeng, and Y.~Pan, ``Emoface: Audio-driven emotional 3d face animation,'' in \emph{2024 IEEE Conference Virtual Reality and 3D User Interfaces (VR)}.\hskip 1em plus 0.5em minus 0.4em\relax IEEE, 2024, pp. 387--397.

\bibitem{nocentini2025emovoca}
F.~Nocentini, C.~Ferrari, and S.~Berretti, ``Emovoca: Speech-driven emotional 3d talking heads,'' in \emph{2025 IEEE/CVF Winter Conference on Applications of Computer Vision (WACV)}.\hskip 1em plus 0.5em minus 0.4em\relax IEEE, 2025, pp. 2859--2868.

\bibitem{wang2020learning}
Z.~Wang, P.~Yu, Y.~Zhao, R.~Zhang, Y.~Zhou, J.~Yuan, and C.~Chen, ``Learning diverse stochastic human-action generators by learning smooth latent transitions,'' in \emph{Proceedings of the AAAI conference on artificial intelligence}, vol.~34, no.~07, 2020, pp. 12\,281--12\,288.

\bibitem{guo2020action2motion}
C.~Guo, X.~Zuo, S.~Wang, S.~Zou, Q.~Sun, A.~Deng, M.~Gong, and L.~Cheng, ``Action2motion: Conditioned generation of 3d human motions,'' in \emph{Proceedings of the 28th ACM international conference on multimedia}, 2020, pp. 2021--2029.

\bibitem{martinez2017human}
J.~Martinez, M.~J. Black, and J.~Romero, ``On human motion prediction using recurrent neural networks,'' in \emph{Proceedings of the IEEE conference on computer vision and pattern recognition}, 2017, pp. 2891--2900.

\bibitem{petrovich2021action}
M.~Petrovich, M.~J. Black, and G.~Varol, ``Action-conditioned 3d human motion synthesis with transformer vae,'' in \emph{Proceedings of the IEEE/CVF international conference on computer vision}, 2021, pp. 10\,985--10\,995.

\bibitem{petrovich2022temos}
------, ``Temos: Generating diverse human motions from textual descriptions,'' in \emph{European Conference on Computer Vision}.\hskip 1em plus 0.5em minus 0.4em\relax Springer, 2022, pp. 480--497.

\bibitem{gong2023tm2d}
K.~Gong, D.~Lian, H.~Chang, C.~Guo, Z.~Jiang, X.~Zuo, M.~B. Mi, and X.~Wang, ``Tm2d: Bimodality driven 3d dance generation via music-text integration,'' in \emph{Proceedings of the IEEE/CVF International Conference on Computer Vision}, 2023, pp. 9942--9952.

\bibitem{tevet2022human}
G.~Tevet, S.~Raab, B.~Gordon, Y.~Shafir, D.~Cohen-Or, and A.~H. Bermano, ``Human motion diffusion model,'' \emph{arXiv preprint arXiv:2209.14916}, 2022.

\bibitem{shafir2023human}
Y.~Shafir, G.~Tevet, R.~Kapon, and A.~H. Bermano, ``Human motion diffusion as a generative prior,'' \emph{arXiv preprint arXiv:2303.01418}, 2023.

\bibitem{chen2024text}
X.~Chen, ``Text-driven human motion generation with motion masked diffusion model,'' \emph{arXiv preprint arXiv:2409.19686}, 2024.

\bibitem{zhang2024motiondiffuse}
M.~Zhang, Z.~Cai, L.~Pan, F.~Hong, X.~Guo, L.~Yang, and Z.~Liu, ``Motiondiffuse: Text-driven human motion generation with diffusion model,'' \emph{IEEE transactions on pattern analysis and machine intelligence}, vol.~46, no.~6, pp. 4115--4128, 2024.

\bibitem{ma2023talkclip}
Y.~Ma, S.~Wang, Y.~Ding, B.~Ma, T.~Lv, C.~Fan, Z.~Hu, Z.~Deng, and X.~Yu, ``Talkclip: Talking head generation with text-guided expressive speaking styles,'' \emph{arXiv preprint arXiv:2304.00334}, 2023.

\bibitem{sun2024avi}
Y.~Sun, W.~Chu, H.~Zhou, K.~Wang, and H.~Koike, ``Avi-talking: Learning audio-visual instructions for expressive 3d talking face generation,'' \emph{IEEE Access}, vol.~12, pp. 57\,288--57\,301, 2024.

\bibitem{wu2024mmhead}
S.~Wu, Y.~Li, Y.~Yan, H.~Duan, Z.~Liu, and G.~Zhai, ``Mmhead: Towards fine-grained multi-modal 3d facial animation,'' in \emph{Proceedings of the 32nd ACM International Conference on Multimedia}, 2024, pp. 7966--7975.

\bibitem{ravdess}
S.~R. Livingstone and F.~A. Russo, ``The ryerson audio-visual database of emotional speech and song (ravdess): A dynamic, multimodal set of facial and vocal expressions in north american english,'' \emph{PloS one}, vol.~13, no.~5, p. e0196391, 2018.

\bibitem{iwase2020song2face}
S.~Iwase, T.~Kato, S.~Yamaguchi, T.~Yukitaka, and S.~Morishima, ``Song2face: Synthesizing singing facial animation from audio,'' in \emph{SIGGRAPH Asia 2020 Technical Communications}, 2020, pp. 1--4.

\bibitem{flame}
\BIBentryALTinterwordspacing
T.~Li, T.~Bolkart, M.~J. Black, H.~Li, and J.~Romero, ``Learning a model of facial shape and expression from {4D} scans,'' \emph{ACM Transactions on Graphics, (Proc. SIGGRAPH Asia)}, vol.~36, no.~6, pp. 194:1--194:17, 2017. [Online]. Available: \url{https://doi.org/10.1145/3130800.3130813}
\BIBentrySTDinterwordspacing

\bibitem{danvevcek2022emoca}
R.~Dan{\v{e}}{\v{c}}ek, M.~J. Black, and T.~Bolkart, ``Emoca: Emotion driven monocular face capture and animation,'' in \emph{Proceedings of the IEEE/CVF Conference on Computer Vision and Pattern Recognition}, 2022, pp. 20\,311--20\,322.

\bibitem{feng2021learning}
Y.~Feng, H.~Feng, M.~J. Black, and T.~Bolkart, ``Learning an animatable detailed 3d face model from in-the-wild images,'' \emph{ACM Transactions on Graphics (ToG)}, vol.~40, no.~4, pp. 1--13, 2021.

\bibitem{filntisis2022visual}
P.~P. Filntisis, G.~Retsinas, F.~Paraperas-Papantoniou, A.~Katsamanis, A.~Roussos, and P.~Maragos, ``Visual speech-aware perceptual 3d facial expression reconstruction from videos,'' \emph{arXiv preprint arXiv:2207.11094}, 2022.

\bibitem{ekman1978facial}
P.~Ekman and W.~V. Friesen, ``Facial action coding system,'' \emph{Environmental Psychology \& Nonverbal Behavior}, 1978.

\bibitem{hakanpaa2019emotion}
T.~Hakanp{\"a}{\"a}, T.~Waaramaa, and A.-M. Laukkanen, ``Emotion recognition from singing voices using contemporary commercial music and classical styles,'' \emph{Journal of Voice}, vol.~33, no.~4, pp. 501--509, 2019.

\bibitem{scherer2017expression}
K.~R. Scherer, J.~Sundberg, B.~Fantini, S.~Trznadel, and F.~Eyben, ``The expression of emotion in the singing voice: Acoustic patterns in vocal performance,'' \emph{The Journal of the Acoustical Society of America}, vol. 142, no.~4, pp. 1805--1815, 2017.

\bibitem{wu2024speechcuellm}
Z.~Wu, Z.~Gong, L.~Ai, P.~Shi, K.~Donbekci, and J.~Hirschberg, ``Beyond silent letters: Amplifying llms in emotion recognition with vocal nuances,'' \emph{arXiv preprint arXiv:2407.21315}, 2024.

\bibitem{lewis2020retrieval}
P.~Lewis, E.~Perez, A.~Piktus, F.~Petroni, V.~Karpukhin, N.~Goyal, H.~K{\"u}ttler, M.~Lewis, W.-t. Yih, T.~Rockt{\"a}schel \emph{et~al.}, ``Retrieval-augmented generation for knowledge-intensive nlp tasks,'' \emph{Advances in neural information processing systems}, vol.~33, pp. 9459--9474, 2020.

\bibitem{wei2022chain}
J.~Wei, X.~Wang, D.~Schuurmans, M.~Bosma, F.~Xia, E.~Chi, Q.~V. Le, D.~Zhou \emph{et~al.}, ``Chain-of-thought prompting elicits reasoning in large language models,'' \emph{Advances in neural information processing systems}, vol.~35, pp. 24\,824--24\,837, 2022.

\bibitem{cohn2007observer}
J.~F. Cohn, Z.~Ambadar, and P.~Ekman, ``Observer-based measurement of facial expression with the facial action coding system,'' \emph{The handbook of emotion elicitation and assessment}, vol.~1, no.~3, pp. 203--221, 2007.

\bibitem{radford2023robust}
A.~Radford, J.~W. Kim, T.~Xu, G.~Brockman, C.~McLeavey, and I.~Sutskever, ``Robust speech recognition via large-scale weak supervision,'' in \emph{International conference on machine learning}.\hskip 1em plus 0.5em minus 0.4em\relax PMLR, 2023, pp. 28\,492--28\,518.

\bibitem{ho2020denoising}
J.~Ho, A.~Jain, and P.~Abbeel, ``Denoising diffusion probabilistic models,'' \emph{Advances in neural information processing systems}, vol.~33, pp. 6840--6851, 2020.

\bibitem{radford2021learning}
A.~Radford, J.~W. Kim, C.~Hallacy, A.~Ramesh, G.~Goh, S.~Agarwal, G.~Sastry, A.~Askell, P.~Mishkin, J.~Clark \emph{et~al.}, ``Learning transferable visual models from natural language supervision,'' in \emph{International conference on machine learning}.\hskip 1em plus 0.5em minus 0.4em\relax PmLR, 2021, pp. 8748--8763.

\bibitem{peebles2023scalable}
W.~Peebles and S.~Xie, ``Scalable diffusion models with transformers,'' in \emph{Proceedings of the IEEE/CVF international conference on computer vision}, 2023, pp. 4195--4205.

\bibitem{thambiraja2023imitator}
B.~Thambiraja, I.~Habibie, S.~Aliakbarian, D.~Cosker, C.~Theobalt, and J.~Thies, ``Imitator: Personalized speech-driven 3d facial animation,'' in \emph{Proceedings of the IEEE/CVF international conference on computer vision}, 2023, pp. 20\,621--20\,631.

\bibitem{kim2025deeptalk}
J.~Kim, J.~Cho, J.~Park, S.~Hwang, D.~E. Kim, G.~Kim, and Y.~Yu, ``Deeptalk: Dynamic emotion embedding for probabilistic speech-driven 3d face animation,'' in \emph{Proceedings of the AAAI Conference on Artificial Intelligence}, vol.~39, no.~4, 2025, pp. 4275--4283.

\bibitem{comanici2025gemini}
G.~Comanici, E.~Bieber, M.~Schaekermann, I.~Pasupat, N.~Sachdeva, I.~Dhillon, M.~Blistein, O.~Ram, D.~Zhang, E.~Rosen \emph{et~al.}, ``Gemini 2.5: Pushing the frontier with advanced reasoning, multimodality, long context, and next generation agentic capabilities,'' \emph{arXiv preprint arXiv:2507.06261}, 2025.

\bibitem{xie2024adan}
X.~Xie, P.~Zhou, H.~Li, Z.~Lin, and S.~Yan, ``Adan: Adaptive nesterov momentum algorithm for faster optimizing deep models,'' \emph{IEEE Transactions on Pattern Analysis and Machine Intelligence}, vol.~46, no.~12, pp. 9508--9520, 2024.

\bibitem{huang2024beat}
Z.~Huang, X.~Xu, C.~Xu, H.~Zhang, C.~Zheng, J.~Qin, and S.~He, ``Beat-it: Beat-synchronized multi-condition 3d dance generation,'' in \emph{European conference on computer vision}.\hskip 1em plus 0.5em minus 0.4em\relax Springer, 2024, pp. 273--290.

\bibitem{siyao2022bailando}
L.~Siyao, W.~Yu, T.~Gu, C.~Lin, Q.~Wang, C.~Qian, C.~C. Loy, and Z.~Liu, ``Bailando: 3d dance generation by actor-critic gpt with choreographic memory,'' in \emph{Proceedings of the IEEE/CVF Conference on Computer Vision and Pattern Recognition}, 2022, pp. 11\,050--11\,059.

\bibitem{tseng2023edge}
J.~Tseng, R.~Castellon, and K.~Liu, ``Edge: Editable dance generation from music,'' in \emph{Proceedings of the IEEE/CVF conference on computer vision and pattern recognition}, 2023, pp. 448--458.

\end{thebibliography}



%



%

\begin{IEEEbiography}[{\includegraphics[width=1in,height=1.25in,clip,keepaspectratio]{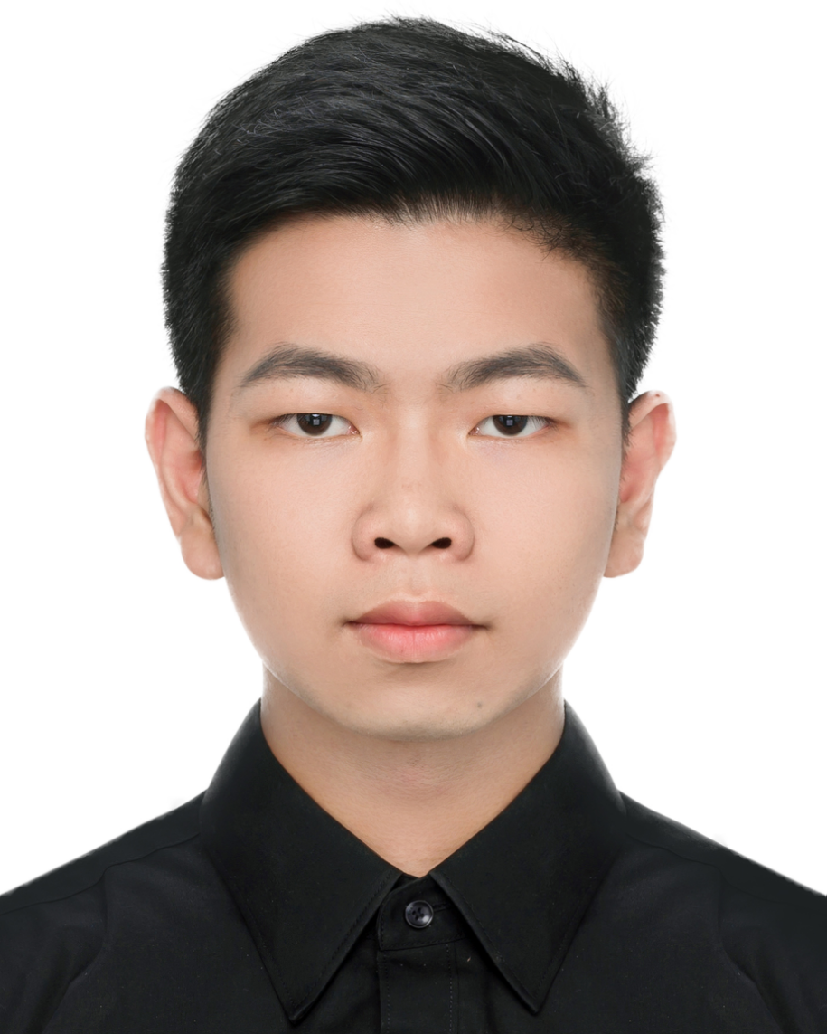}}]{Zikai Huang} is currently pursuing a Ph.D. degree in the School of Computer Science and Engineering, South China University of Technology. His research interests include computer vision, computer graphics and multimodal learning.
\end{IEEEbiography}

\begin{IEEEbiography}[{\includegraphics[width=1in,height=1.25in,clip,keepaspectratio]{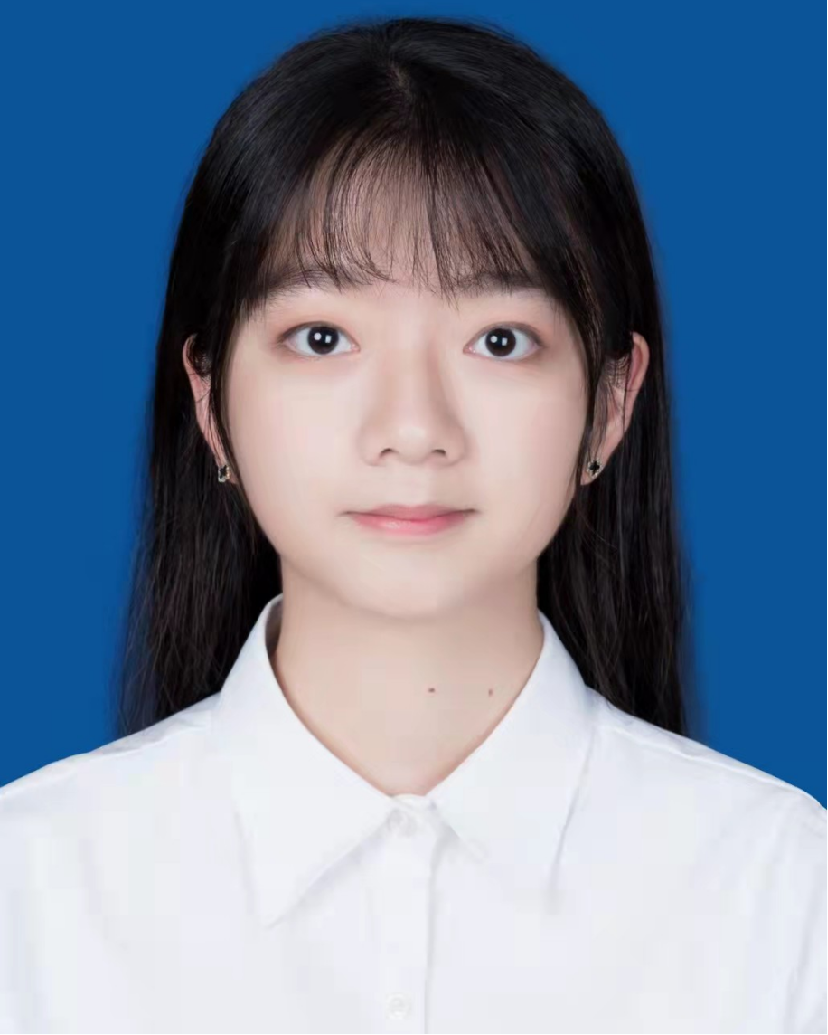}}]{Yihan Zhou} is is a first-year M.S. student in the School of Computer Science and Engineering, South China University of Technology. Her research interests include computer vision, computer graphics and multimodal learning.
\end{IEEEbiography}

\begin{IEEEbiography}[{\includegraphics[width=1in,height=1.25in,clip,keepaspectratio]{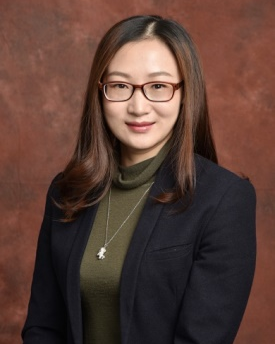}}]{Xuemiao Xu} received the BS and MS degrees in computer science and engineering from South China University of Technology, in 2002 and 2005, respectively, and the PhD degree in computer science and engineering from The Chinese University of Hong Kong in 2009. She is currently a professor with the School of Computer Science and Engineering, South China University of Technology. Her research interests include object detection, tracking, recognition, and image, video understanding and synthesis, particularly their applications in the intelligent transportation.
\end{IEEEbiography}

\begin{IEEEbiography}[{\includegraphics[width=1in,height=1.25in,clip,keepaspectratio]{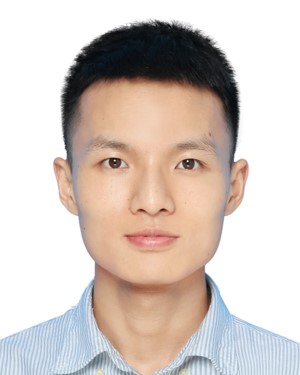}}]{Cheng Xu} received his Ph.D. degree in computer science and technology from South China University of Technology, China, in 2023. He is currently a Post-Doctoral Fellow at The Hong Kong Polytechnic University. His research interests include computer vision, generative models, and medical image analysis.
\end{IEEEbiography}

\begin{IEEEbiography}[{\includegraphics[width=1in,height=1.25in,clip,keepaspectratio]{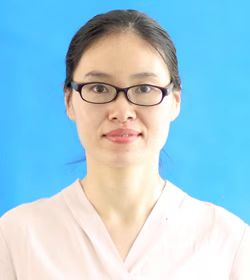}}]{Xiaofen Xing} (Member, IEEE) received the B.S., M.S., and Ph.D. degrees from the South China University of Technology, Guangzhou, China, in 2001, 2004, and 2013, respectively. Since 2017, she has been an Associate Professor with the School of Electronic and Information Engineering, South China University of Technology. Her main research interests include speech emotion analysis, image/video processing, and human-computer interaction.
\end{IEEEbiography}

\begin{IEEEbiography}[{\includegraphics[width=1in,height=1.25in,clip,keepaspectratio]{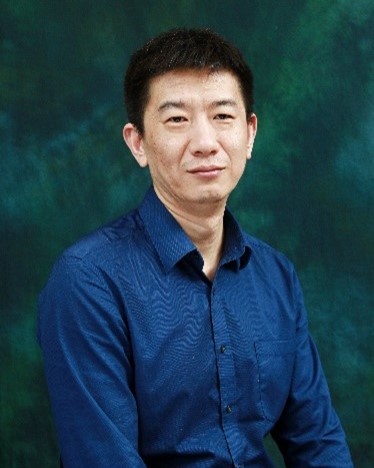}}]{Jing Qin} (Senior Member, IEEE) is a professor and the director of The Centre for Smart Health, School of Nursing, The Hong Kong Polytechnic University. He also serves as the director of CAS-Hong Kong Joint Laboratory for Multimodal Medical Molecular Imaging, and the director of the Program of Master of Science in Health Informatics, The Hong Kong Polytechnic University. His research interests are innovatively harnessing advanced artificial intelligence (AI) and extended reality (XR) techniques in various medicine and healthcare applications. Prof. Qin has won the 2024 CES (Consumer Technology Associate) 2024 Innovation Awards, The Higher Education Outstanding Scientific Research Output Awards (Science and Technology) to Chinese Ministry of Education (Second Prize) in 2022, the 2019 MICCAI Young Scientist Publication Impact Award, the 2017 Medical Image Analysis-MICCAI'17 Best Paper Award.
\end{IEEEbiography}

\begin{IEEEbiography}[{\includegraphics[width=1in,height=1.25in,clip,keepaspectratio]{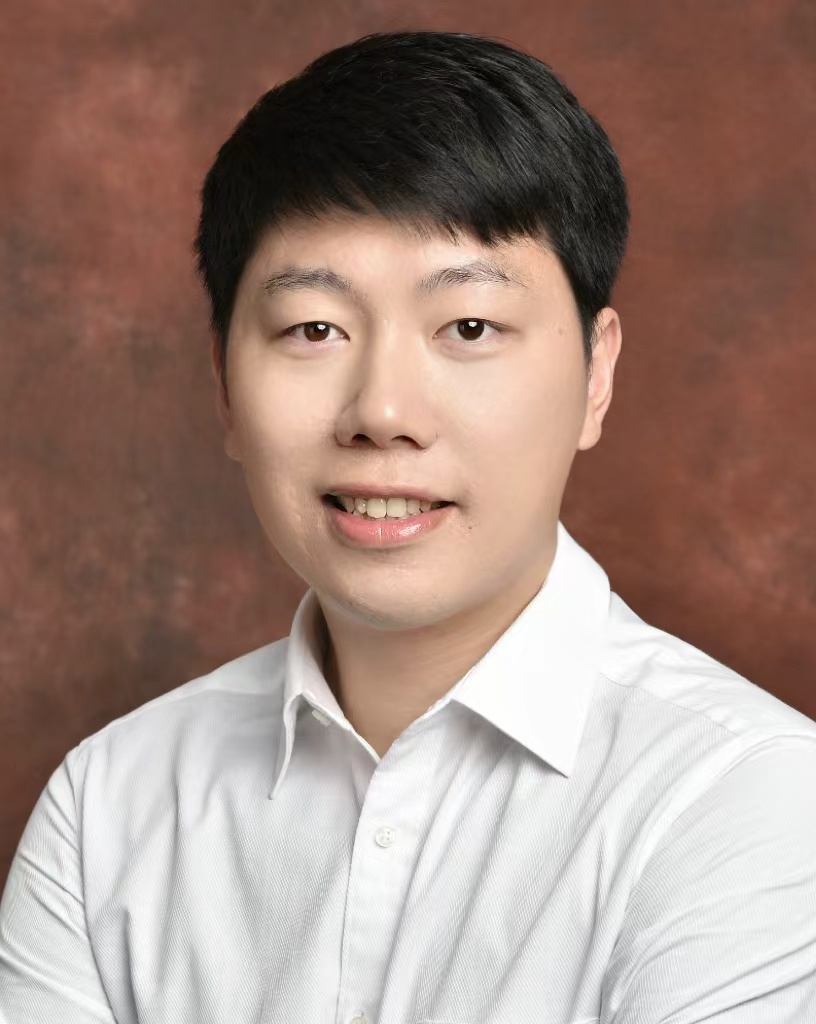}}]{Shengfeng He} (Senior Member, IEEE) is an associate professor in the School of Computing and Information Systems at Singapore Management University. Previously, he was a faculty member at South China University of Technology (2016--2022). He earned his B.Sc. and M.Sc. from Macau University of Science and Technology (2009, 2011) and a Ph.D. from City University of Hong Kong (2015). His research focuses on computer vision and generative models. He has received awards including the Google Research Award, PerCom 2024 Best Paper Award, and the Lee Kong Chian Fellowship. He is a senior IEEE member and distinguished CCF member. He serves as lead guest editor for IJCV and associate editor for IEEE TNNLS, IEEE TCSVT, Visual Intelligence, and Neurocomputing. He is an area chair/senior PC member for NeurIPS, ICLR, ICML, AAAI, IJCAI, and BMVC, and will serve as Conference Chair of Pacific Graphics 2026.
\end{IEEEbiography}




\end{document}


\sloppy

%
\title{Think2Sing: Orchestrating Structured Motion Subtitles for Singing-Driven 3D Head Animation\\ - Supplementary Materials -}

\author{
	Zikai~Huang,
  Yihan~Zhou,
	Xuemiao~Xu,
	Cheng~Xu,
	Xiaofen~Xing,~\IEEEmembership{Member,~IEEE}, \\
	Jing~Qin,~\IEEEmembership{Senior Member,~IEEE},
	and Shengfeng~He,~\IEEEmembership{Senior Member,~IEEE}
}


%
%

\markboth{IEEE Transactions on Visualization and Computer Graphics}%
{Huang \MakeLowercase{\textit{et al.}}: Think2Sing}
%




\makeatletter
\long\def\@IEEEtitleabstractindextextbox#1{\parbox{0.922\textwidth}{#1}}
\makeatother

\maketitle

\IEEEdisplaynontitleabstractindextext

%
\IEEEpeerreviewmaketitle

This supplementary document provides additional details to complement the main paper.
\refsec{dataset} describes the implementation of motion subtitle annotations for key facial regions, including the eyebrows, mouth, eyes, and neck pose.
\refsec{prompt_for_singcot} presents the full prompt design for Sing-CoT.
\refsec{application} illustrates the practical use of motion subtitles, highlighting their role as both semantic guidance for generation and as a human-interpretable editing interface.
Together, these sections offer a comprehensive view of the dataset construction, prompting strategy, and application potential of our proposed framework.
\section{Motion Subtitle Annotation} \label{sec:dataset}
Implementation details of the motion subtitle annotations for the eyebrows, mouth, eyes, and neck pose are provided as follows:

\textbf{Eyebrows and Mouth.}
The region-wise velocity signal is constructed by selecting vertices within the target region and computing their frame-wise displacement.
For each region $r$ and frame $i$, the velocity is computed as:
\begin{equation}
vel_{r,i} = \frac{1}{|V_r|} \sum_{v \in V_r} \lVert v_i - v_{i-1} \rVert_2,
\end{equation}
where $v_i$ denotes the 3D position of vertex $v$ at frame $i$, and $V_r$ is the set of vertices in region $r$.
The velocity curve $vel_{r,i}$ captures local motion dynamics, where local minima often indicate motion slowdowns and potential state transitions, while local maxima correspond to peak motion changes.
Detection is performed in four steps:

1) local minima in the velocity curve are identified as candidate boundary points;

2) a pair of successive minima frames $i_1$ and $i_2$ is retained if the AU intensity difference exceeds a predefined threshold $\tau$, thereby filtering out spurious minima unrelated to meaningful expression changes;

3) within the interval $(i_1, i_2)$, the frame with maximum velocity is designated as the expression onset, representing the most pronounced motion;

\begin{table}[t]
  \noindent\begin{minipage}[t]{.48\textwidth}
    \centering
    \captionof{table}{
        Description vocabulary for eyebrows and mouth.
    }
    \begin{tabular}{c|c} \toprule \hline
        \textbf{Motion}                & \textbf{\texttt{<des>}}                                                               \\ \hline
        Eyebrow lower                  & ``lower'', ``furrow'', ``pull down''      \\
        Eyebrow raise                  & ``lift'', ``arch'', ``raise''      \\
        Smile during singing           & ``smile'', ``smile with parted lips''      \\
        Smile without singing          & ``soft-smile'', ``smile with closed lips''   \\
        Mouthfrown                     & ``frown'', ``downturn mouth''   \\ \hline
        \textbf{Intensity}             & \textbf{\texttt{<adv>}}                                                                \\ \hline
        Strongly                       & \makecell{``sharply'', ``heavily'', ``strongly'', \\``significantly'', ``extremely''}       \\
        Slightly                       & \makecell{``gently'', ``slightly'', \\``mildly'', ``faintly'', ``a little''}      \\  \hline \bottomrule
    \end{tabular}
    \label{brow_mouth_vocabulary}
\end{minipage}
\end{table}

\begin{table}[t]
    \caption{Description vocabulary for the eyes motion.}
    \centering
        \centering
        \begin{tabular}{c|c} \toprule \hline
            \textbf{Eyes Motion} & \textbf{\texttt{<des>}}     \\ \hline
            Eye widen            & ``open wide'', ``widen''    \\
            Eye squint           & ``squint'', ``half-closed'' \\
            Eye close            & ``shut'', ``close''         \\ \hline \bottomrule
        \end{tabular}
    \label{eye_vocabulary}
\end{table}

\begin{table}[t]
    \caption{Description vocabulary for the neck motion.}
    \centering
    \begin{minipage}[t]{1\columnwidth}
        \centering
        \begin{tabular}{c|c} \hline
            \textbf{Neck Motion} & \textbf{\texttt{<des>}}                        \\ \hline
            Head up              & ``raise'', ``lift''                            \\
            Head down            & ``lower'', ``drop''                            \\
            Head turn            & ``turn left/right'', ``rotate left/right''     \\
            Head tilt            & ``tilt left/right'', ``lean left/right''       \\
            Head shake           & ``sway to the rhythm'', ``shake side to side'' \\ \hline
        \end{tabular}
    \end{minipage}
    \label{neck_vocabulary}
\end{table} 

4) a subsequent minimum frame $i_3$ after $i_2$ is selected if the AU intensity difference between $i_2$ and $i_3$ exceeds $\tau$, with the frame of maximum velocity in $(i_2, i_3)$ defining the expression offset.
The onset and offset frames thus obtained specify the annotated expression interval.
This procedure balances temporal precision and semantic validity.
Velocity extrema provide robust temporal cues for motion segmentation, while AU intensity differences ensure that segmented intervals reflect substantive facial expression changes rather than noise.
For the eyebrows, we focus on brow lower~(AU4) and brow raiser~(AU1).
For the mouth, we consider lip corner puller~(AU12) and lip corner depressor~(AU15), which are linked to smiling and frowning expressions, respectively.
Based on the normalized AU intensity difference between frames $i_1$ and $i_2$, a difference greater than 0.25 is regarded as a slight change in intensity, whereas a difference greater than 0.5 is regarded as a strong change.
We list the vocabulary for eyebrows and mouth descriptions in \reftab{brow_mouth_vocabulary}.

\begin{figure*}[t]
\centering
\includegraphics[width=\linewidth,scale=1.00]{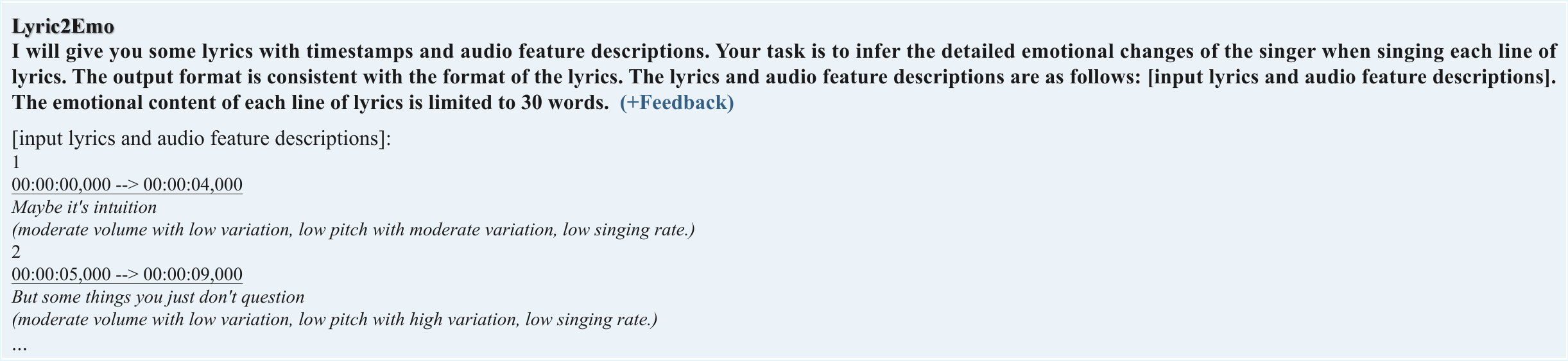}
\caption{
Prompt template for the Lyrics2Emo step.
} 
\label{lyrics2emo}	
\end{figure*}

\begin{figure*}[t]
\centering
\includegraphics[width=\linewidth,scale=1.00]{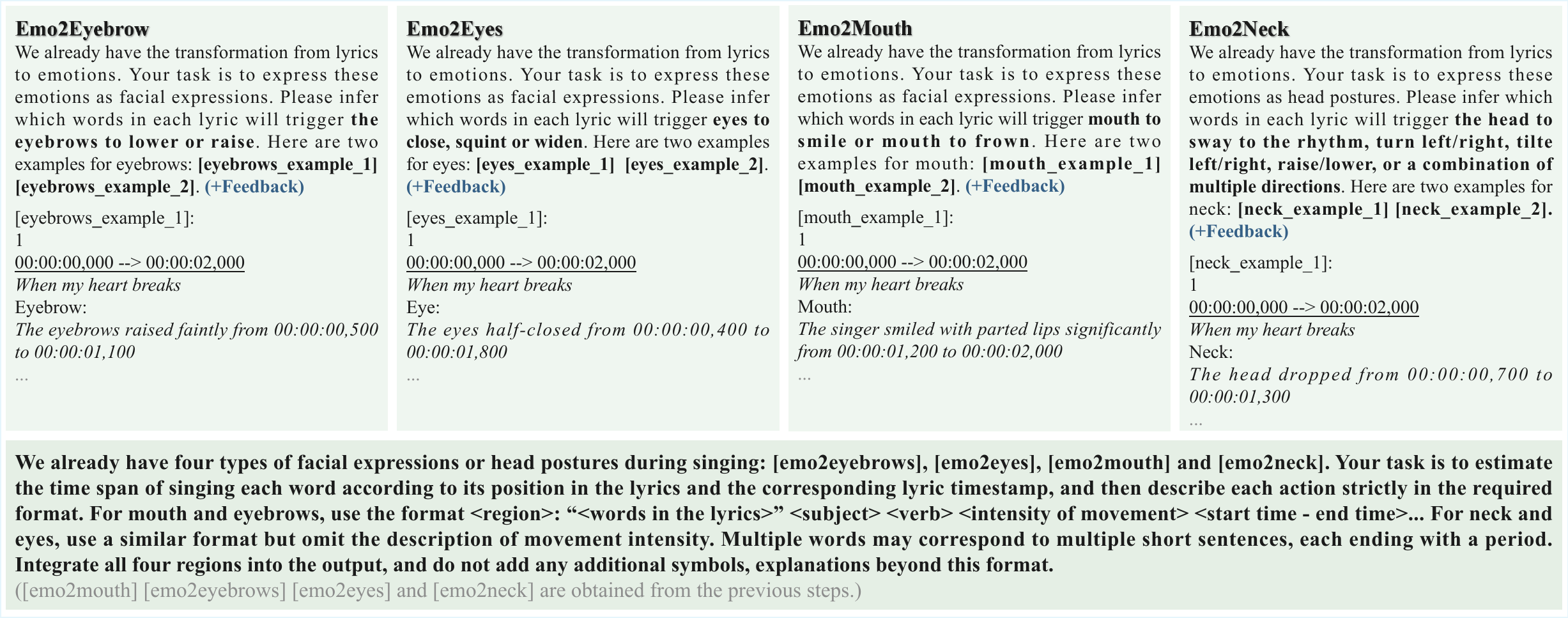}
\caption{Prompt template for the Emo2Subs step.} 
\label{emo2subs}	
\end{figure*}

\begin{figure*}[t]
\centering
\includegraphics[width=\linewidth,scale=1.00]{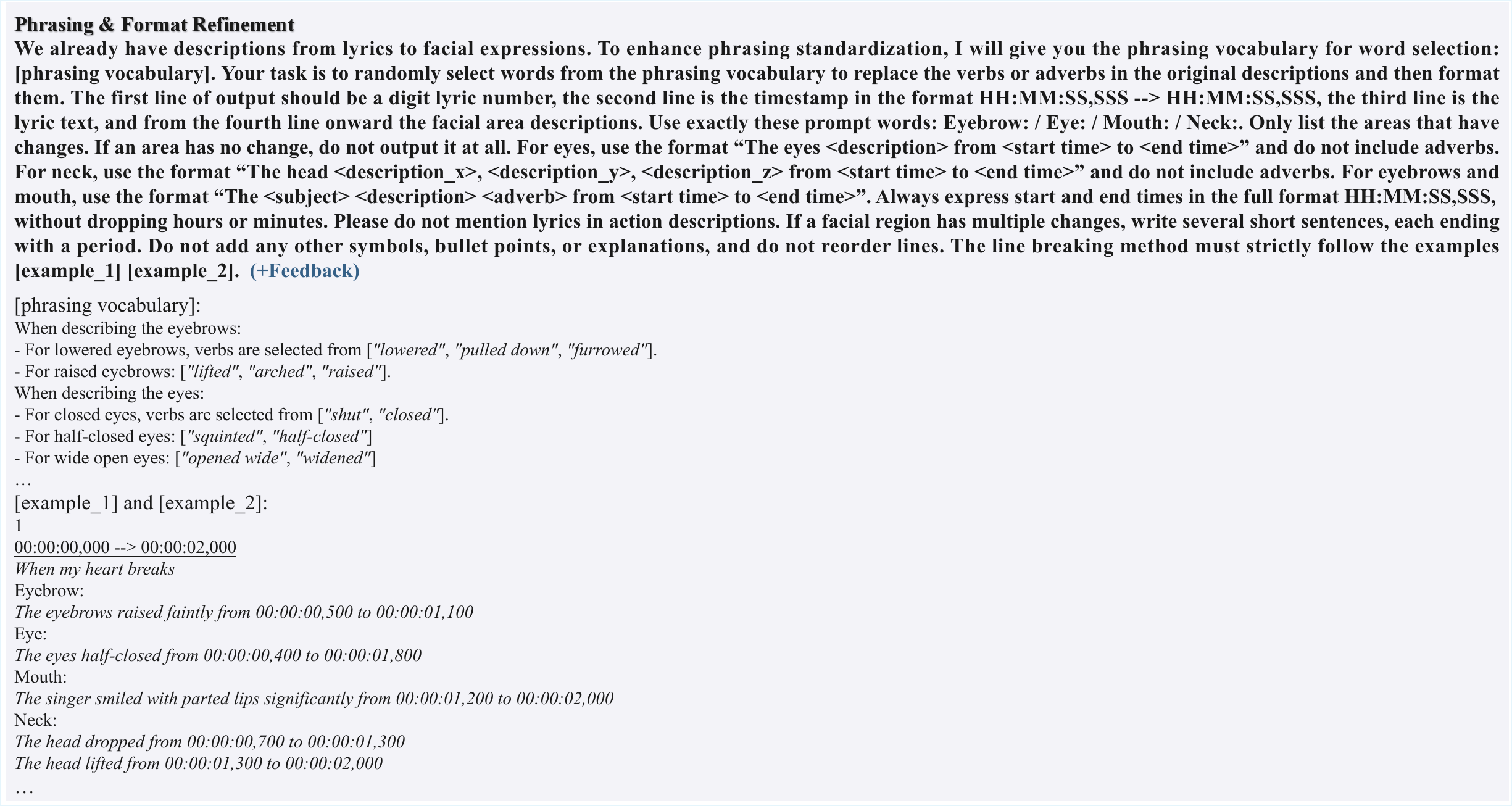}
\caption{Prompt template for the Phrasing \& Format Refinement step.\vspace{-3mm}}
\label{phrasing_format}	
\end{figure*}

\begin{figure*}[t]
\centering
\includegraphics[width=\linewidth,scale=1.00]{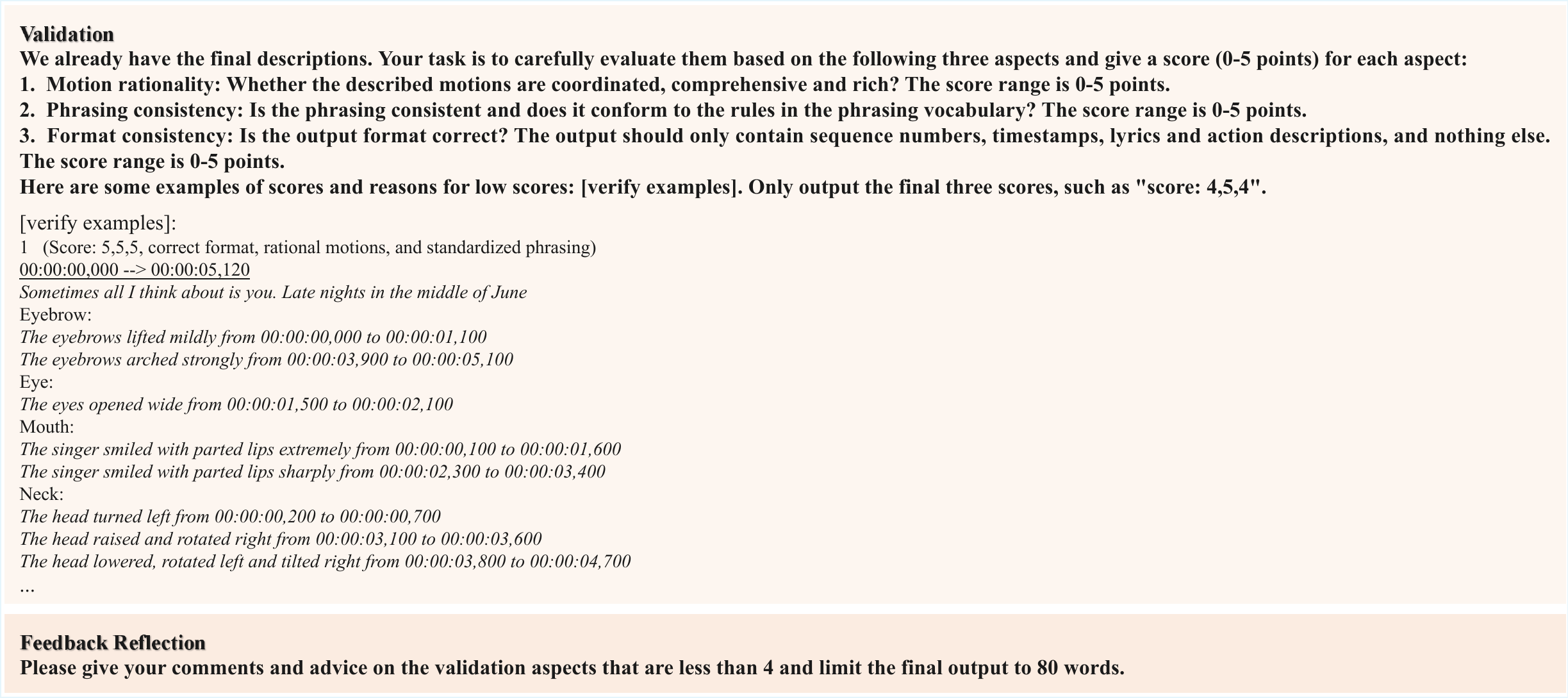}
\caption{Prompt template for the Validation and Feedback Reflection step.}
\label{validation_feedback}	
\end{figure*}

\textbf{Eyes.}
The average distance between the upper and lower eyelid landmarks in a neutral gaze is approximately 9.5 mm.
Eye motion is described as ``widen'' when the distance exceeds this value, ``squint'' when it falls between 4 mm and 6 mm, and ``close'' when it is less than 4 mm.
To eliminate brief blinks, we discard eye movements that last less than 0.5 seconds. The vocabulary for eye motion descriptions is provided in \reftab{eye_vocabulary}.

\textbf{Neck Pose.}
For neck pose, movements with a rotation angle of less than 10 degrees are filtered out to exclude minor, non-significant motions.
After the neck motion sequences are consolidated, actions with a duration shorter than 0.5 seconds are discarded to eliminate brief, potentially irrelevant movements.
The vocabulary for neck pose descriptions is provided in \reftab{neck_vocabulary}.

\section{Prompt for Sing-CoT} \label{sec:prompt_for_singcot}
We provide the detailed prompt for Sing-CoT in this section, which consists of the following parts:

\textbf{Lyrics2Emo.}
The prompt takes lyrics, audio description and revision feedback from the Validation step as input.
This design emphasizes a multimodal perspective, since relying on lyrics alone would neglect the prosodic and acoustic cues that are essential for capturing the intended emotional expression in singing.
By conditioning the prompt on both modalities, the model is encouraged to produce emotion labels that are musically grounded as well as semantically faithful to the lyrics.
The detailed prompt template is shown in~\reffig{lyrics2emo}.

\textbf{Emo2Subs.}
Once the emotion is determined, the next step generates motion subtitles aligned with the lyrics.
This step leverages exemplar guidance and emotion conditioning to ensure the outputs are both expressive and stylistically consistent.
Feedback from validation is incorporated in later rounds, enabling iterative improvement and correction.
The detailed prompt template is shown in~\reffig{emo2subs}.

\textbf{Phrasing \& Format Refinement.}
The refinement step introduces a dual-focus mechanism designed to enhance both the expressiveness and the structural reliability of the generated motion subtitles.
At the semantic level, the model is guided by a curated vocabulary that aligns the phrasing more closely with the training domain, thereby mitigating stylistic drift and promoting consistency with domain-specific usage.
At the structural level, explicit formatting constraints are enforced to ensure conformity with the subtitle schema required for subsequent generation.
The detailed prompt template is shown in~\reffig{phrasing_format}.

\textbf{Validation \& Feedback Reflection.}
The final step introduces a validation module that evaluates each set of motion subtitles in terms of motion rationality, phrasing consistency, and format consistency.
Rather than serving only as a quality check, this step plays a central role in our framework by converting evaluation results into structured feedback that is reintegrated into earlier steps.
Through this reflection loop, the system transcends one-shot generation and engages in iterative reasoning with self-correction.
This step functions as an active mechanism for producing outputs that are increasingly coherent, expressive, and reliable.
The detailed prompt template is shown in~\reffig{validation_feedback}.

\section{Application} \label{sec:application}
Motion subtitles provide clear semantic guidance for generation and can also function as a human-interpretable editing interface.
This feature enhances user interaction and offers greater flexibility in content creation.
By specifying or modifying motion subtitles, our method enables fine-grained head animation editing.
To demonstrate this capability, we present several examples in the \href{https://zikaihuangscut.github.io/Think2Sing/}{\textcolor{blue}{project page}}, showcasing the versatility of our approach in generating diverse head movements based on user-defined motion subtitles.


%





\ifCLASSOPTIONcaptionsoff
  \newpage
\fi